\documentclass[manuscript]{aastex631}

\usepackage{graphicx}	
\usepackage{amsmath}
\usepackage{amssymb}	
\usepackage{longtable}
\usepackage{threeparttable}
\usepackage{multirow}
\usepackage{booktabs}
\usepackage{array}
\usepackage[figuresright]{rotating}
\usepackage[T1]{fontenc}  
\usepackage{colortbl}    
\usepackage{amsmath}

\begin{document}

\title{Towards Understanding the Origin of \textit{Swift} Gamma-Ray Bursts Driven by Magnetars}

\author{C. T. Hao}
\affiliation{Department of Physics, College of Physics, Guizhou Univesity, Guiyang, 550025, China\\}

\author{J. H. Jing}
\affiliation{Department of Physics, College of Physics, Guizhou Univesity, Guiyang, 550025, China\\}
\affiliation{These authors contributed to the article equally\\}
\author{X. L. Han}
\affiliation{Department of Physics and Astronomy, Butler University, Indianapolis, IN 46208, USA\\}
\affiliation{These authors contributed to the article equally\\}
\author{H. R. Lan}
\affiliation{School of Electrical and Electronic Engineering,
Nanyang Technological University, 50 Nanyang Avenue, Singapore\\}
\affiliation{These authors contributed to the article equally\\}
\author{W. C. Du}
\affiliation{Engineering Laboratory for Optoelectronic Technology and Advanced Manufacturing,\\ School of Physics, Henan Normal University, Xinxiang, 453007, China\\}
\author{X. N. Liu}
\affiliation{Department of Physics, College of Physics, Guizhou Univesity, Guiyang, 550025, China\\}
\author{Z. B. Zhang$^{\star}$}
\affiliation{Department of Physics, College of Physics, Guizhou Univesity, Guiyang, 550025, China\\}
\email{zbzhang@gzu.edu.cn}

\author{H. C. Liu$^{\ast}$}
\affiliation{Engineering Laboratory for Optoelectronic Technology and Advanced Manufacturing,\\ School of Physics, Henan Normal University, Xinxiang, 453007, China\\}
\email{hcliu12@sina.cn}
\author{J. F. Wu}
\affiliation{Department of Physics, College of Physics, Guizhou Univesity, Guiyang, 550025, China\\}
\author{X. L. Xia}
\affiliation{Department of Physics, College of Physics, Guizhou Univesity, Guiyang, 550025, China\\}

\collaboration{20}{(AAS Journals Data Editors)}



\begin{abstract}
	
In this paper, we analyze a sample of\textit{ Swift} gamma-ray bursts (GRBs) with extended emissions in $\gamma$-rays and/or X-ray plateaus that may be driven by magnetars. Multi-wavelength data and multi-standards have been adopted to investigate the issue jointly. First, we find that GRBs with both extended emission and X-ray plateau satisfy a three-parameter relation between the luminosity and the end time of X-ray plateaus and the $\gamma$-ray isotropic energy as  $L_X\varpropto T_a^{-1.13}E_{\gamma,iso}^{0.74}$, which is consistent with that of normal GRBs. Second, we distinguish these GRBs in the plane of magnetic field versus period of neutron star and find that almost all GRBs but GRB 211024B have reasonable periods and majority of them could be powered by magnetars. Third, we standardize the X-ray afterglows with distinct characteristics and find that the standard X-ray light curves with/without plateaus are significantly different. The standardized X-ray plateaus are similar to the mean temporal profile of magnetars. Fourth, it is verified with a K-S test that all types of GRBs except short ones have the similar distributions of redshift and isotropic energy in the observer/rest frame. GRBs with internal plateaus are significantly different from those of normal long GRBs and GRBs with external plateaus and/or extended emissions. Interestingly, the isotropic energy distributions of GRBs with internal and external plateaus are identical with those of short and long GRBs, respectively. Overall, our study can bring solid evidence that the fascinating magnetars could have multi-formation channels to account for not only short but also long GRBs with either internal or external X-ray plateaus as well.

\end{abstract}



	\section{Introduction} 
Gamma-ray bursts (GRBs) are among the universe's most intense yet brief flashes of light. These phenomena generally arise from collisions within highly relativistic jets of material \citep[][for a review]{1992MNRAS.258P..41R,1986ApJ...308L..43P,2018pgrb.book.....Z}. The interactions between these jets and the interstellar medium will produce afterglow, observable across various low-energy wavelengths \citep{1976PhFl...19.1130B}. GRBs are traditionally classified based on their duration, using the $T_{90}$ parameter: bursts with durations shorter than 2 seconds are classified as short gamma-ray bursts (SGRBs), while those lasting longer than 2 seconds are categorized as long gamma-ray bursts (LGRBs) \citep{1981Ap&SS..80....3M,1993ApJ...413L.101K,2008A&A...484..293Z}.

LGRBs are primarily associated with the core-collapse of massive stars \citep{1998Natur.395..670G,2003Natur.423..847H,2013ApJ...776...98X}. This collapse can result in the formation of either a black hole surrounded by an accretion disk or a rapidly rotating, highly magnetized neutron star known as a magnetar \citep{2012ApJ...757...69U,2014ApJ...783...10S}. The magnetar scenario has gained increasing support as the favored explanation \citep{2015JHEAp...7...64B}, positioning magnetars as a likely central engine driving LGRBs \citep{2011MNRAS.413.2031M,2014MNRAS.443...67M,2018MNRAS.480.4402L}. This hypothesis is reinforced by the frequent association of LGRBs with supernovae, a connection first noted with GRB 980425 \citep{1998Natur.395..670G} and further validated by subsequent studies \citep{2004ApJ...609L...5M,2003ApJ...599..394M,2023arXiv231014310H}. Magnetars have also been proposed as the primary energy source behind the luminous SN-associated GRBs \citep{2010ApJ...717..245K,2014MNRAS.443...67M}.

\cite{2007MNRAS.382.1029K} studied the behavior of outflows from a magnetar interacting with an inwardly collapsing stellar progenitor, analyzing the dynamics for 200 ms following core collapse. Their results indicated a non-relativistic, jet-like outflow with collimation characteristics. Further research by \cite{2008MNRAS.383L..25B} demonstrated that magnetar winds, produced after the core collapse of a massive star, can form an axial jet. These findings collectively support the potential of millisecond magnetars as central engines for certain LGRBs, offering a consistent model for these energetic phenomena.

 SGRBs, on the other hand, are thought to originate from the mergers of compact binary systems, such as binary neutron stars (BNS) \citep{1986ApJ...308L..43P,1989Natur.340..126E} or neutron star-black hole (NS-BH) systems \citep{1991AcA....41..257P}. The detection of the gravitational wave (GW) event GW170817, alongside SGRB 170817A by the Gamma-ray Burst Monitor (GBM) on the \textit{Fermi} spacecraft, provided strong evidence for the BNS merger origin of this SGRB \citep{2017ApJ...848L..13A}. This discovery solidified the link between BNS mergers and SGRBs. Previous studies \citep{2011PhRvD..83l4008H,2012PTEP.2012aA304S,2013PhRvD..87b4001H} have shown that the merger of a BNS system can produce either a massive neutron star or a black hole. The outcome is influenced by the total mass of the binary system post-merger and the equation of state of the neutron stars \citep{2014PhRvD..89d7302L}. If the remnant is sufficiently massive, it may collapse quickly into a black hole or form an unstable supermassive neutron star. A less massive remnant may instead become a supermassive proto-magnetar \citep{2013PhRvD..87b4001H,2002MNRAS.334..481R,2013ApJ...771L..26G}.
 
  Real-time data analysis by the \citet{2019GCN.25324....1L} identified GW S190814bv as a candidate signal from the merger of compact binary objects. Subsequent analysis suggested that these objects were likely a neutron star (NS) and a black hole (BH) \citep{2019GCN.25333....1L}. Observations of excess components in the infrared or optical afterglows of SGRBs further support a binary merger origin \citep{2013ApJ...774L..23B,2016EPJWC.10908002J,2021arXiv210907694J}. The excess is generally interpreted as emission from a radioactive kilonova, produced by the rapid neutron capture process (r-process) in unstable heavy elements undergoing radioactive decay \citep{1998ApJ...507L..59L}. The merger dynamics between the BNS and the NS-BH systems differ notably. In the NS-BH mergers, the neutron star is gradually disrupted and accreted by the black hole, leading to significant mass transfer and the formation of a disk substantial enough to potentially power a GRB \citep{2005MNRAS.356...54D}. Although the NS-BH mergers have not been observed to produce magnetars, they can still give rise to kilonova events \citep{2011ApJ...730..141Z}. 

Observations from the \textit{Swift} satellite indicate that about one-quarter of SGRBs exhibit extended X-ray emissions lasting between 10 and 100 seconds with a fluence larger than that of the initial bursts \citep{2008AIPC.1000..280N,2009AJ....138.1690P}. It is found that around 15\% of SGRBs show extended $\gamma$-ray emissions \citep{2006ApJ...643..266N}. These extended emissions (EE) consist of a softer and longer-lasting burst that follows the initial spike, generally with a lower peak flux \citep{2006ApJ...643..266N}. Two primary hypotheses have been proposed to explain its origin. The first one assumes that a spinning-down magnetar serves as the central engine for EE GRBs \citep{2012MNRAS.419.1537B,2013MNRAS.431.1745G,2008MNRAS.385.1455M}, while the second one is attributed to the magnetar wind activity \citep{2012MNRAS.419.1537B}. Additionally, \cite{2014MNRAS.438..240G} proposed that the EE might result from accretion onto a magnetar from a surrounding disk. The EE components typically last around 100 seconds since the trigger time in the rest frame, which challenges the conventional BNS merger model and instead suggests that a rapidly rotating, highly magnetized neutron star is a likely central engine for the EE of SGRBs. Of course, there are studies of extended emission within the context of black hole engines \citep[e.g.][]{2015ApJ...804L..16K}.

Swift satellite aslo detected  approximately 15\% of LGRBs  \citep{1995ApJ...452..145K,2009A&A...505..569B,2008ApJ...685L..19B} and 10\% of SGRBs with precursor radiations before the trigger time \citep{2010ApJ...723.1711T}. Moreover, about 80\% of LGRBs and 5\% of SGRBs behave a plateau structure in their X-ray light curves  \citep{2009MNRAS.397.1177E,2013MNRAS.430.1061R,2014A&A...565A..72M}. Several theories have been proposed to explain the origin of the X-ray plateaus. For example,  \cite{2023A&A...675A.117R} pointed out that  continuous energy injection from a long-lived central engine such as the  millisecond pulsars or magnetars by the dipole radiation can contribute to the X-ray plateaus. The process of receding accretion onto black holes is thought to have an effective impact on the X-ray plateaus \citep{2014MNRAS.438..240G,2023MNRAS.522.5848L,2019ApJS..245....1T}. \cite{2023mgm..conf.3141P} argued that the plateau could be resulted from the GRB jet materials gliding through a surrounding wind-like medium. Interestingly, the fireball model predicted that the X-ray plateau may arise from the dynamics of the expanding fireball that powers the GRB \citep{1998ApJ...496L...1R}. More interestingly, \cite{2020MNRAS.492.2847B} verified by simulations that the relativistic jets can also produce the X-ray plateaus if only they are viewd off-axis. Additionally, the X-ray plateaus together with a sharp drop in luminosity can be also explained by a hypothesis of BH \citep{2008MNRAS.388.1729K,2009ApJ...700.1047C,2011ApJ...734...35C,2025JHEAp..4700384L}, or interaction of the jet with circumburst medium \citep{2014MNRAS.445.2414V,2020ApJ...900..193D}.


Theoretical models propose that these plateau phases may result from energy injections powered by magnetars \citep{1998A&A...333L..87D,2001ApJ...552L..35Z,2006MNRAS.372L..19F,2007ApJ...665..599T}. Depending on whether the magnetar's spin-down energy influences the forward shock wave or dissipates internally before reaching the shock front, the plateau can be classified as either ``external'' or ``internal'' \citep{2006ApJ...642..354Z,2010MNRAS.402..705L,2015JHEAp...7...64B}. After the plateau phase, the sharp decline in the X-ray light curve is consistent with the magnetic dipole spin-down of a protomagnetar, potentially culminating in its collapse into a black hole \citep{2013MNRAS.430.1061R,2015ApJ...805...89L,2008MNRAS.385L..10T}. This sharp drop in luminosity challenges the hypothesis of a black hole instead of a rapidly spinning millisecond magnetar as the central engine. Observationally, magnetars are thought to be the predominant central engines powering both LGRBs and SGRBs \citep{2015JHEAp...7...64B}. However, how to understand the diversity or similarity of magnetar-like (ML) GRBs is still an open question. In particular, \cite{2020ApJ...895...58G} poited out that the SGRBs with EE behave distinct energies and galactic offsets in contrast with other SGRBs.

The structure of this paper is arranged as follows. Section 2 gives the GRB sample selection. In Section 3, we apply the $L-T-E$ \citep{2019ApJS..245....1T}  and $B-P$ \citep{2018ApJ...869..155S} relationships to identify the most probable magnetar central engines. Section 4 provides a statistical analysis of $T_{90}$, $E_{\gamma,iso}$, and redshift ($z$) for GRBs across different categories. Section 5 presents the key findings of our study. The implications and challenges of our main results are shown in Section 6. Throughout this paper, the cosmological parameters of ${H_0} = 70 \ \rm km\,{\rm s^{ - 1}}\,{\rm Mpc^{ - 1}}$, ${\Omega _M} = 0.27$ and ${\Omega _\Lambda }=0.73$ have been adopted.

\section{Sample Selection} \label{sec:2}

To deeply investigate how SGRBs and LGRBs with/without EE components as magnetar candiates differ from other types of GRBs, we first choose 39 SGRBs from \cite{2020ApJ...895...58G} and 42 EE GRBs from \cite{2020RAA....20..201Z} and some additional GRBs with redshift reported on the \textit{Swift }official website as of March 21, 2025 to constitute our total GRB sample, of which 163 GRBs including 36 SGRBs and 127 LGRBs have been selected and listed in table \ref{tab:1}, in which 75\% and 92\% of bursts exhibit precursors and EEs, respectively, and 103 GRBs with both precursors, main peaks, and EEs. 158 out of 163 GRBs have the observed X-ray afterglows, among which 109 GRBs including 92 LGRBs and 17 SGRBs exhibit X-ray plateaus, corresponding to percentages of 72\% and 47\%, respectively. Notably, the fraction of SGRBs with EE in \cite{2020ApJ...895...58G} can go up to 62\% if the criterion of $S/N>3$ for the EE judgement is employed as in \cite{2020RAA....20..201Z}. Column 1 lists the GRB name, followed by the redshift in Column 2. Column 3 presents the duration of $T_{90}$ reported on the \textit{Swift} table. Column 4 indicates whether a GRB has precursor emission. Column 5 specifies whether the main burst consists of a single or multiple pulses. Column 6 denotes whether the EE component has been detected. Column 7 presents the plateau type of X-ray afterglows (see Section 3.1 for details). Column 8 lists the prompt $\gamma$-ray fluence in\textit{ Swift} BAT energy band. Columns 9 and 10 present the k-correction and isotropic $\gamma$-ray energy of each GRB, respectively. Note that the intrinsic duration is defined by ${t_{int}} = {T_{{\rm{90}}}}/(1 + z)$. Following \cite{ 2018PASP..130e4202Z}, we calculate the isotropic $\gamma$-ray energy $E_{\gamma,\rm iso}$ after the  k-correction for frequency changes from the observer frame to the rest frame.

\section{central engine}

Although BHs and magnetars are two prospestive candidates, the central engine of GRBs still remains mysterious. If magnetars power GRBs, we would expect to observe some specific signatures such as the plateau plus sharp decay in X-ray light curves, the EE component, and the precursor of both SGRBs and LGRBs. In the following, we will test the consistency of magnetars with the empirical relations of $L-T-E$ and $B-P$ by the linear regression technique.

\subsection{X-ray afterglow plateau}

The X-ray afterglow data have been taken from the official \textit{Swift}/XRT website\footnote{\url{https://www.swift.ac.uk/xrt_curves/}}. It should be emphasized that the X-ray plateaus contain at least three data points and the declining X-ray afterglows in a power-law form possess at least two data points in order to obtain the reliable fitting. Besides, 17 GRBs with rare observations have been taken out. As a result, 146 GRBs are left for the subsequent analysis of X-ray afterglows.

To get the basic parameters of X-ray light curves, we have fitted 109 well-sampled X-ray data by using the smoothly broken power-law function \citep[e.g.][]{2019ApJS..245....1T,2023MNRAS.tmp.1633L} as
\begin{equation}
	\label{4}
	{F_X}\left( t \right) = {\rm{ }}{F_{X0}}{\left[ {{{\left( {\frac{t}{{{\rm{ }}{T_0}}}} \right)}^{{\alpha _1}\omega }} + {{\left( {\frac{t}{{{\rm{ }}{T_0}}}} \right)}^{{\alpha _2}\omega }}} \right]^{ - 1{\rm{ /}}\omega }},
\end{equation}
where $\alpha_1$ and $\alpha_2$ respectively describe the power-law indices of the early plateau and the late-on decay part. $T_0$ is the observed end time of plateaus. ${F_{X0}} \times {2^{ - 1/\omega }}$ is the X-ray flux at the time $T_0$. The smoothness parameter bridging the plateau and decay segments is assumed to be $\omega=3$ in our analysis. For 37 X-ray afterglows without obviously rebrightening phenomena, we simply fit them with a single power-law (SPL) or an injection-free (IF) form. \cite{2024ApJ...960...77D} found that GRBs with the IF X-ray afterglows hold the largest peak energy in comparison to those GRBs with quasi-plateau X-ray afterglows. Regarding the X-ray afterglow fitted by Eq. (\ref{4}), we define that an effective plateau should have a power-law index of $ - 0.75 < {\alpha _1} < 0.75$. Otherwise, the rebrightening X-ray structures can not be treated as plateaus. Based on this criterion, we then classify the X-ray plateaus into internal X-ray plateaus (IXPs) with $\alpha _2\geq3$ and external X-ray plateaus (EXPs) with $\alpha _2 <3$, which is slightly different from  \cite{2020ApJ...895...58G}. The best fitting parameters and the X-ray afterglow types are presented in Table \ref{tab:2}, in which we distinguish 109 X-ray plateaus including 86 EXPs and 9 IXPs, and 37 IF X-ray afterglows. The remaining 14 GRBs with $\alpha_1$ outside of the range of $ - 0.75 < {\alpha _1} < 0.75$ .


We find that 140 out of 146 GRBs in our sample exhibit extended emissions (EE) in prompt $\gamma$-rays. Among these, 105 GRBs with EE were successfully fitted using the plateau model, accounting for 66\% (105/146) of GRBs with a plateau phase being associated with EE. Specifically, out of the 86 EXPs, 83 GRBs have the EE components. Notably, 77 out of 86 EXPs belong to LGRBs, 8 out of 9 GRBs with IXPs have the EE components and 6 out of 9 IXPs belong to SGRBs. This indicates a propensity for detecting X-ray light curves with plateau phases in GRBs that exhibit EE, a higher likelihood of encountering EXPs in LGRBs with EE, and larger possibility in identifying IXPs in SGRBs with EE.

\subsection{$L-T-E$ correlation}

\begin{figure}[b]
	\centering
	\includegraphics[width=12cm]{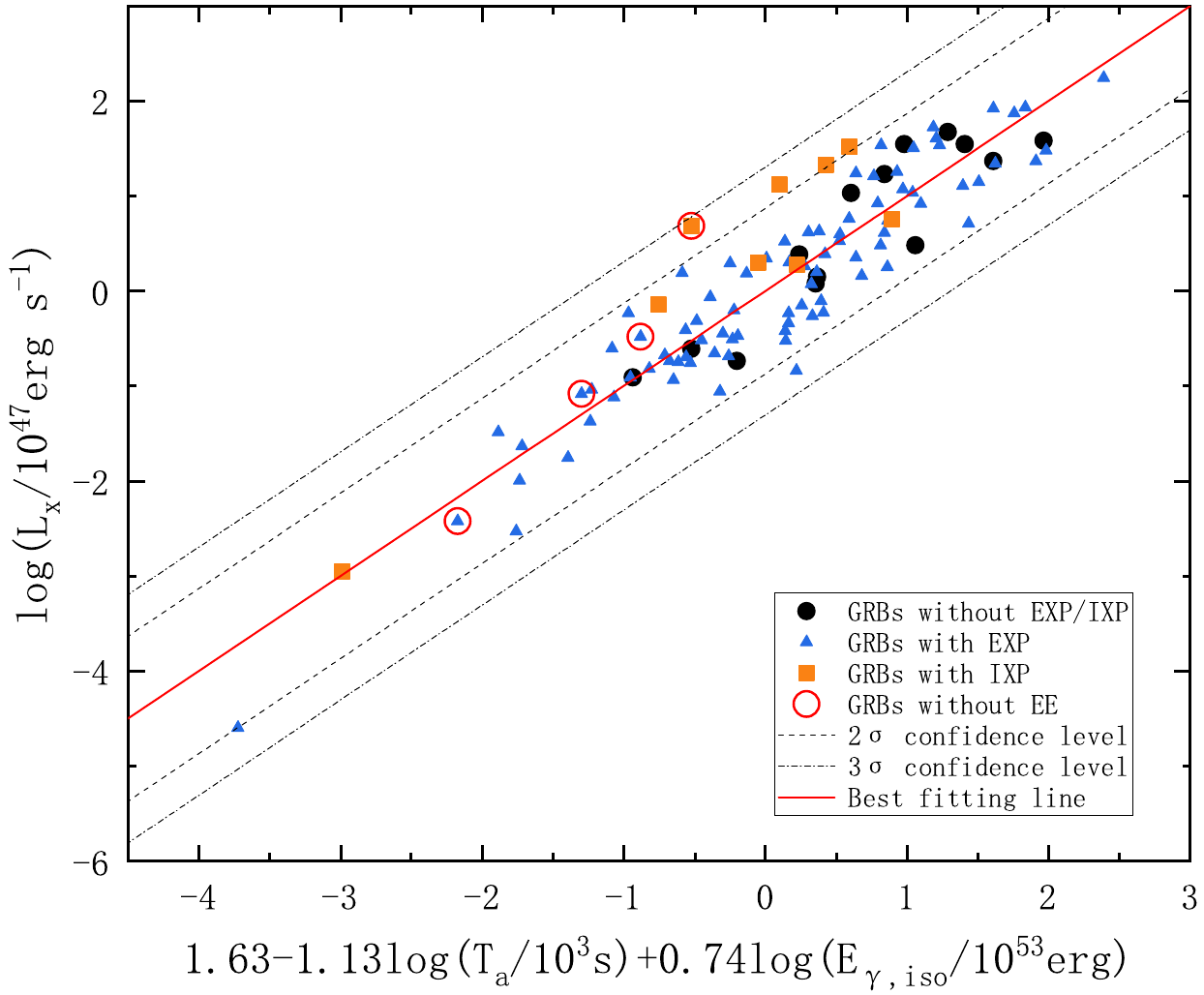}
	\caption{The $L-T-E$ relation of 109 GRBs with well-measured X-ray afterglows. The solid, dashed and dash-dotted lines correspond to the best fit, 2$\sigma$  and 3$\sigma$ confidence levels. All symbols are illustrated in the insert.}
	\label{fig:2}
\end{figure}

\cite{2008MNRAS.391L..79D} identified a significant inverse correlation between the rest-frame end time, $T_{a}=T_0/(1+z)$, of X-ray plateaus and the corresponding luminosity ($L_{X}$ ). The $L_{X}-T_{a}$ correlation was subsequently confirmed by \cite{2010ApJ...722L.215D} and may be biased by selection effects and evolution-independent \citep{2013MNRAS.436...82D,2015MNRAS.451.3898D}. \cite{2012A&A...538A.134X} and \cite{2019ApJS..245....1T} found that both $T_{a}$ and $L_{X}$ were also correlated with $E_{\gamma,iso}$, that is the so-called three-parameter relation ($L-T-E$), which supports the magnetar as the central engine of GRBs with X-ray plateaus.
Here, we verify whether the 109 special GRBs with X-ray plateaus hold the similar three-parameter relation. The X-ray luminosity at the end of plateau is given by
\begin{equation}
	\label{11}
	{{{L}}_{{{X}}}}{\rm{ = 4}}\pi {{D}}_{\rm{L}}^2{{F_X}}{\left( {1 + z} \right)^{ - \left( {2 - \Gamma } \right)}} \ \  (\rm erg\ s^{-1}),
\end{equation}
where $F_X$ is the X-ray flux at the observer-frame end time $T_0$, $z$ is the redshift of a GRB, $\Gamma$ is the power-law index of GRB spectrum. The luminosity distance, $D_{\rm L}$, is computed with
\begin{equation}
	\label{2}
	{D_{L} } = {\rm{c}}H_0^{ - 1}(1 + z)\int_0^z {dz} '{\left[ {{{\left( {1 + z'} \right)}^3}{\Omega _{M}} + {\Omega _\Lambda }} \right]^{ - \frac{1}{2}}}.
\end{equation} 
Following \cite{2018PASP..130e4202Z,2021MNRAS.503.3262Z}, we calculate the isotropic energy of prompt $\gamma$-rays with the following formula
\begin{equation}
	\label{1}
	{E_{\gamma ,iso}} = 4\pi D_L^2 S_{\gamma}{\left( {1 + z} \right)^{ - 1}}K_c\ \ (\rm erg),
\end{equation}
where $S_{\gamma}$ is the $\gamma$-ray fluence obtained from \textit{Swift} GRB table\footnote{\url{https://swift.gsfc.nasa.gov/archive/grb_table/index.php}} and $K_c$ is the $k$-correction factor. These derived quantities are presented in Table \ref{tab:3}. We adopt the Markov Chain Monte Carlo (MCMC) method to fit 109 GRBs with good X-ray afterglows in Figure \ref{fig:2} and obtain the three-parameter correlations between $L_X$, $T_a$ and $E_{\gamma,iso}$ as
\begin{equation}
	\label{7}
	\log \left(\frac{{L_X}}{{{10}^{47}}erg{s^{ - 1}}} \right) = (1.63\pm0.08) - (1.13\pm0.06)\log \left(\frac{{T_a}}{{{10}^3}s} \right) + (0.74\pm0.04)\log \left(\frac{{E_{\gamma ,iso}}}{{{10}^{53}}erg} \right),
\end{equation}
which can be roughly written to be $L_X\varpropto T_a^{-1.13}E_{\gamma,iso}^{0.74}$ with an intrinsic scatter of $0.845^{+0.074}_{-0.070}$.
The existence of $L-T-E$ correlation indicates the 109 GRBs are most probably powered by the central engine of magnetars. Meanwhile, we notice that not only the GRBs with EE but also the GRBs with external/internal X-ray plateaus match the three-parameter relation well.

Notably, our $T_a$ power-law index of -1.13 is consistent with that of \cite{2012A&A...538A.134X} and \cite{2019ApJS..245....1T}. In addition, the $E_{\gamma,iso}$ power-law index of 0.74 is also close to their value of $\sim$0.8, which manifests that most of GRBs in our sample could originate from a magnetar. There are three convincing arguments in the literature that some GRBs can be powered by magnetars. 1) Part of the population of potential GRB-born magnetars can be matched with the local population of pulsars\citep{2014MNRAS.443.1779R}; 2) A phenomenon observed in some sources (e.g. GRB061121)showed that magnetar can turn off its accretion by the temporal increase in the magnetic field \citep{2013ApJ...775...67B}; 3) All models of BH driven GRBs require a phase of approximately constant accretion rate, however SGRBs produced by compact merger does not have any way to get matter for this phase (lack of extended stellar envelope), thus the only know process so far which could create a plateau in SGRBs is an NS spin-down \citep{2001ApJ...552L..35Z}. However, a BH central engine can not be fully ruled out in terms of Eq. (\ref{7}) since both BH and magnetar can create such three-parameter  relation if one treats $E_{\gamma,iso}$ as quantity proportional to the total rotational kinetic energy reserve $I(2\pi/P)^2/2$ in case of magnetar or mass in case of BH \citep{2025JHEAp..4700384L}.

\subsection{$B-P$ Relationship}
\cite{2018ApJ...869..155S} proposed an additional diagnostic method using the correlation between magnetic field strength and spin period ($B-P$) to distinguish the  magnetar-driven GRBs. It proves that the $B-P$ diagram is a valuable tool in identifying magnetar candidates as potential central engine of GRBs \citep{2020ApJ...903L..24L,2018ApJ...869..155S,2022NewA...9701889K}. The alignment of GRB candidates within this framework further supports the hypothesis of magnetars as central engines in GRBs, making the $B-P$ diagram an essential tool for understanding the origin of these high-energy phenomena. To obtain the parameters $B$ and $P$ of a given GRB, we need to fit the X-ray lightcurves with the following X-ray luminosity formula

\begin{equation}
	\label{8}
	{{L_X(t) = 4}}\pi D_L^2F_X(t){\left( {1 + z} \right)^{ - \left( {2 - \Gamma } \right)}}{f_B},
\end{equation}
where $f_B$=1-cos$\theta_j$ denots the beaming factor for a jet half-opening angle $\theta_j$ and $\Gamma$ is the spectral index of SPL GRB spectrum. 65 out of 109 GRBs with X-ray plateaus have the known $\theta_j$ values taken from \cite{2024MNRAS.530.2877L, 2021ApJ...908L...2S, 2023ApJ...959...13R, 2023ApJ...949L..32D, 2023ApJ...950...30Z, 2018PASP..130e4202Z}, the remaining GRBs without measued $\theta_j$ are assumed to $\theta_j=10^\circ$ for SGRBs and $\theta_j=5^\circ$ for LGRBs. If the X-ray luminosity is contributed by the spin-down lumnosity of a NS, the observed luminosity in Eq. (\ref{8}) can be fitted by the nether form \citep{2018ApJ...869..155S} as
\begin{equation}
	\label{9}
	{L_{sd}}(t) = \frac{{{E_{spin}}}}{\tau{{{\left[ {1 + (1 - \alpha )\frac{t}{{{\tau}}}} \right]}^{\frac{2-\alpha}{1-\alpha}}}}},
\end{equation}
in which $\alpha=0$ corresponding to the braking index \textbf{$n=3-\alpha$} for an ideal magnetohydrodynamics (MHD), $\tau=\Omega/(2\dot \Omega)$ is the initial spin-down timescale and
$E_{spin} = (1/2)I\Omega^2$ stands for the NS spin energy. Note that $I$ and $\Omega$ are the spin moment of interia and angular speed of NS, respectively. The above fitting process yields the intial spin period, $P={2\pi}/{\Omega}$, and the dipolar surface magnetic field strength, $B$ as listed in Table \ref{tab:3}

\begin{figure}
	\centering
	\includegraphics[width=12cm]{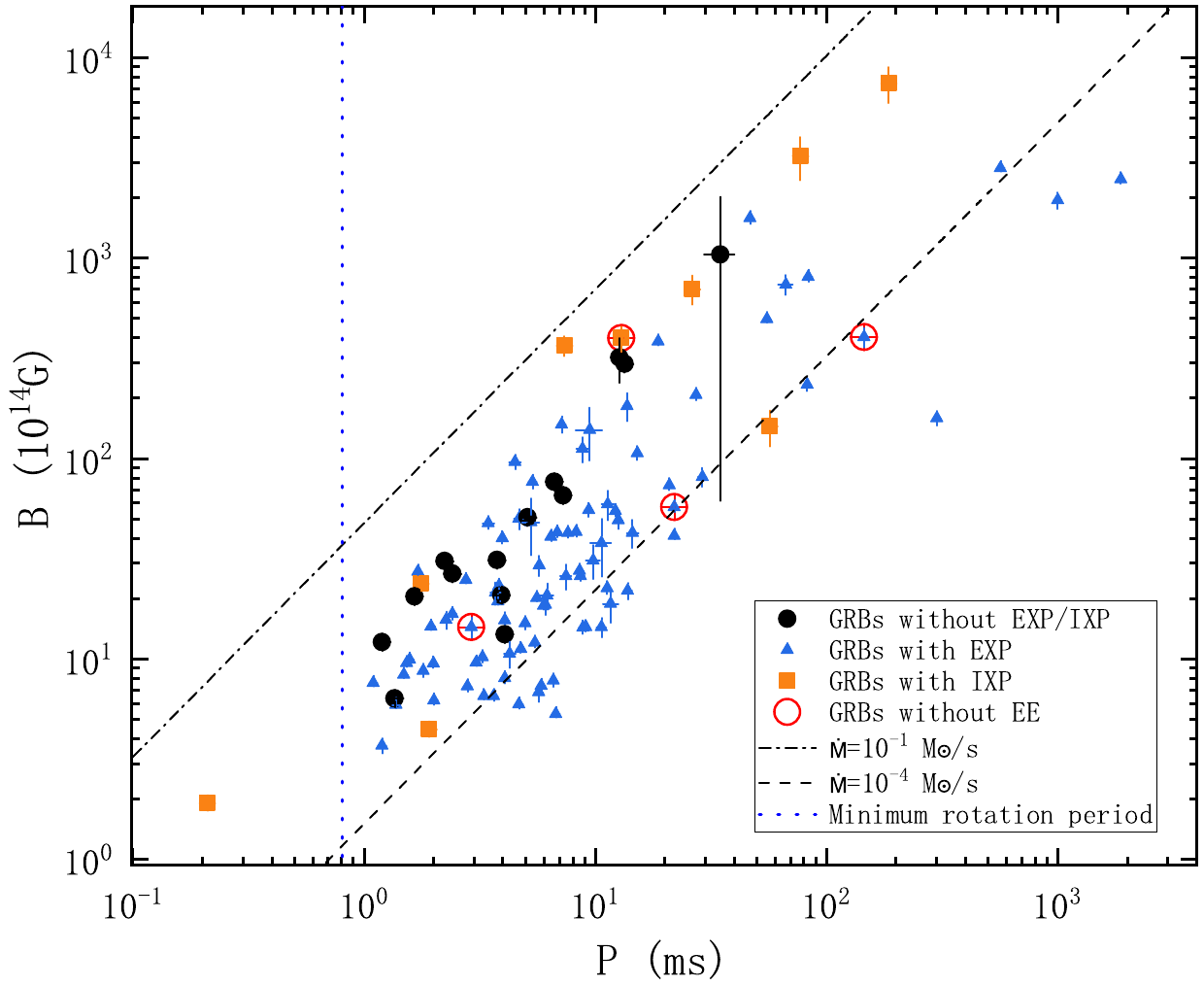}
	\caption{The $B-P$ of 109 GRBs with EEs and/or X-ray plateaus. The dashed and dash-dotted lines indicates the  $B-P$ correlations within the upper and lower mass accretion limits, respectively, The dotted line on the left is the minimum rotation period of 0.81ms.}  
	\label{fig:3}
\end{figure}

On the other hand, the intial spin period and the magnetic field can be also constrained by the following relation \citep{1991PhR...203....1B,2013Ap&SS.346..119P,2018ApJ...869..155S}
\begin{equation}
	\label{BPrelation}
	\dfrac{B}{10^{14} \rm G}\approx15\left(\dfrac{P}{1\  \rm {ms}}\right)^{7/6}\left(\dfrac{\dot M}{0.01\ \rm M_{\odot} \ \rm S^{-1}}\right)^{1/2},
\end{equation}
when two parameters of $R = 12\ km$ and $M = 1.4\ {\rm M_ \odot }$ have been assumed for a typical NS. For a given accretion rate $\dot M$ ranging from ${10^{ - 4}}{\rm M_ \odot }\ {\rm s^{ - 1}} < \dot M < {10^{ - 1}}{\rm M_ \odot }\ {\rm s^{ - 1}}$, we can easily plot two boundary lines in the logarithmic scale between ${P}$ and $B$ as shown in Figure \ref{fig:3}, where those GRBs falling within the bounds are categorized into magnetar-like candidates. Therefore, we can identify from Figure \ref{fig:3} that there are 90 out of 109 GRBs with X-ray plateaus are most probably powered by magnetars. Among the 90 magnetar-like GRBs, 8 IXPs and 68 EXPs are involved and 87 GRBs exhibit the EE components. These results provide the strong evidence that a magnetar should be the central engine of GRBs characterized with not only the IXPs/EXPs but also the EEs.

According to \cite{2014MNRAS.443.1779R}, the minimum allowed rotation period of a NS is given by
\begin{equation}
	\label{77}
	{P_{{\rm{0}}, - 3}} \ge 0.81M_{1.4}^{{{ - 1} \mathord{\left/
				{\vphantom {{ - 1} 2}} \right.
				\kern-\nulldelimiterspace} 2}}R_6^{{3 \mathord{\left/
				{\vphantom {3 2}} \right.
				\kern-\nulldelimiterspace} 2}}\ \ (\rm ms),
\end{equation}
and has been displayed with the vertical dotted line for $R=6\rm $ km and $M=1.4{\rm M_ \odot }$ in Figure \ref{fig:3}, in which we find that only one burst (GRB 211024B) resides within the forbidden zone. It is noticeable that a significant fraction of GRBs in our sample have larger magnetic fields over $10^{16}$ G, which is coincident with some previous estimates \cite[e.g.][]{2018ApJ...869..155S,2022NewA...9701889K}. This implies that the extragalactic magnetars may have larger magnetic fields and total energy outputs than the galactic ones as a whole. The initial magnetic field can be inherited from the progenitor star due to magnetic flux conservation, or the two magnetized NSs in case of merger origin of GRBs. As pointed out by Stratta et al. (2018), for more massive ($M=2.1{\rm M_ \odot }$) and compact ($R\approx10$ km) NS, the maximum spin energy can reach up to $10^{53}$ erg \citep{2018MNRAS.480.1353D} that is obviously larger our means of $E_{\gamma,iso}$ in right panel of Figure \ref{fig:4}. Assuming $M=1.4{\rm M_ \odot }$  to be lower mass limit of a NS, we can only obtain the upper limits of B and P as listed in Table \ref{tab:3}.

\section{Morphology of X-ray afterglows}

\begin{figure}
	\centering
	\includegraphics[width=0.45\columnwidth]{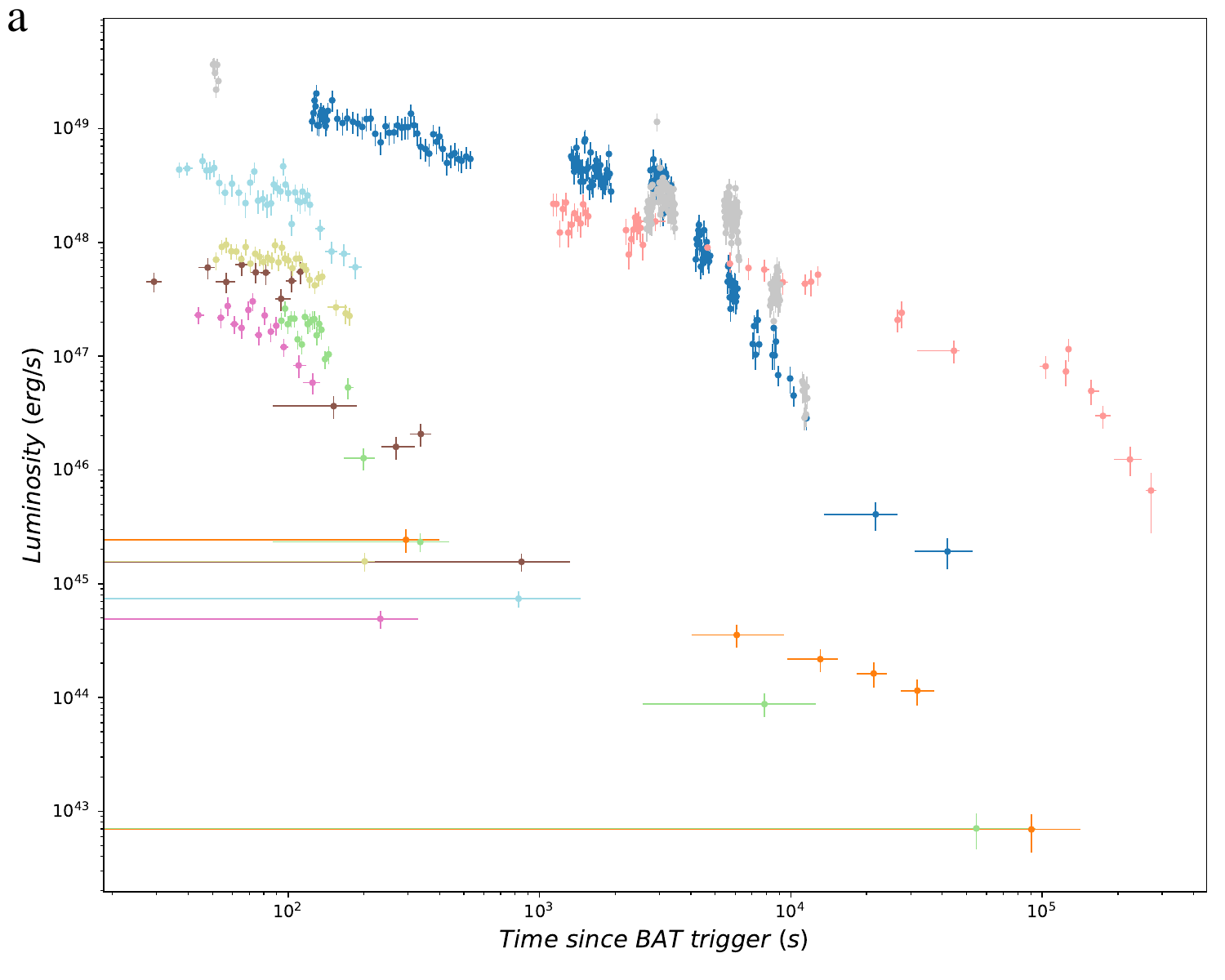}
	\includegraphics[width=0.45\columnwidth]{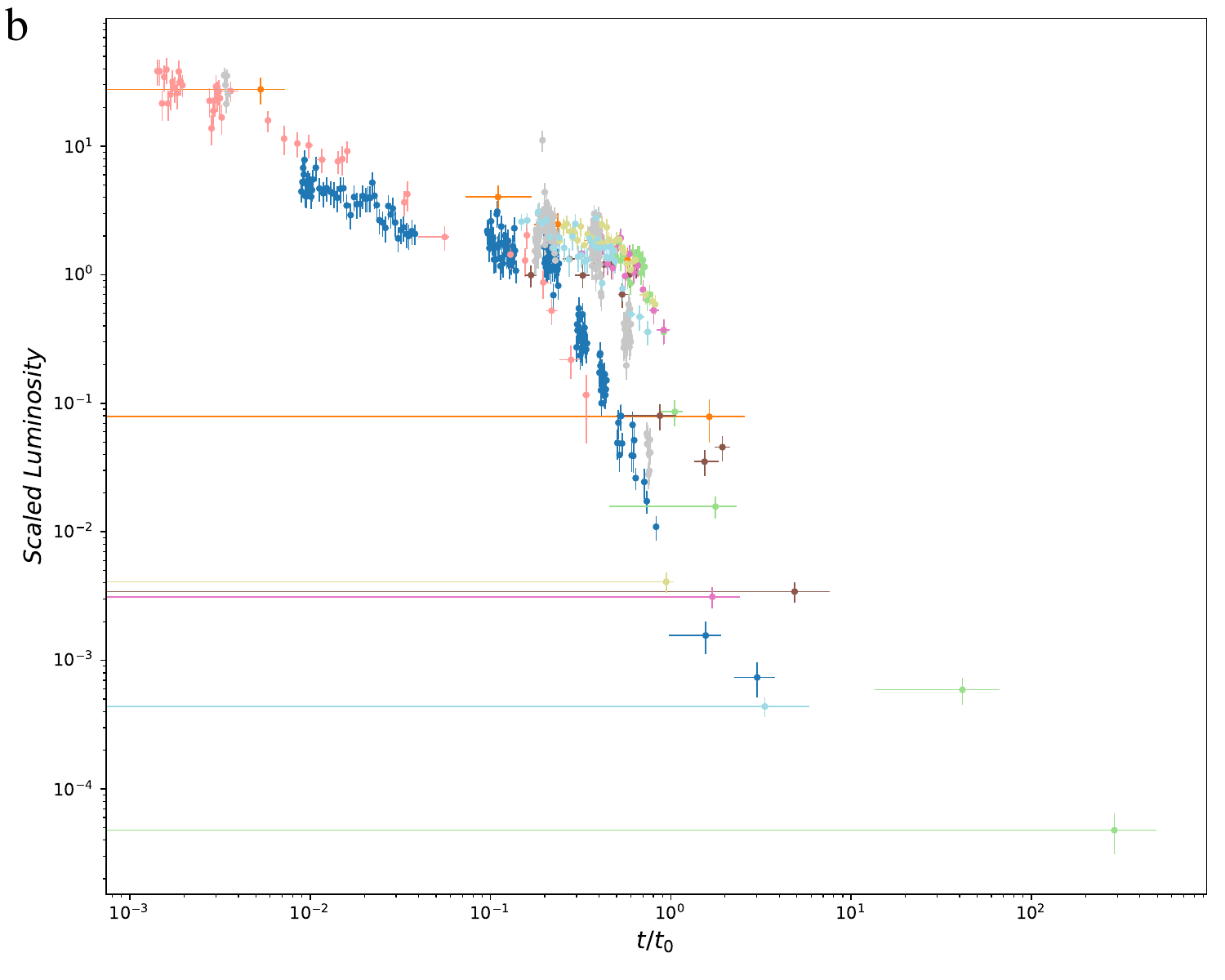}
	\includegraphics[width=0.45\columnwidth]{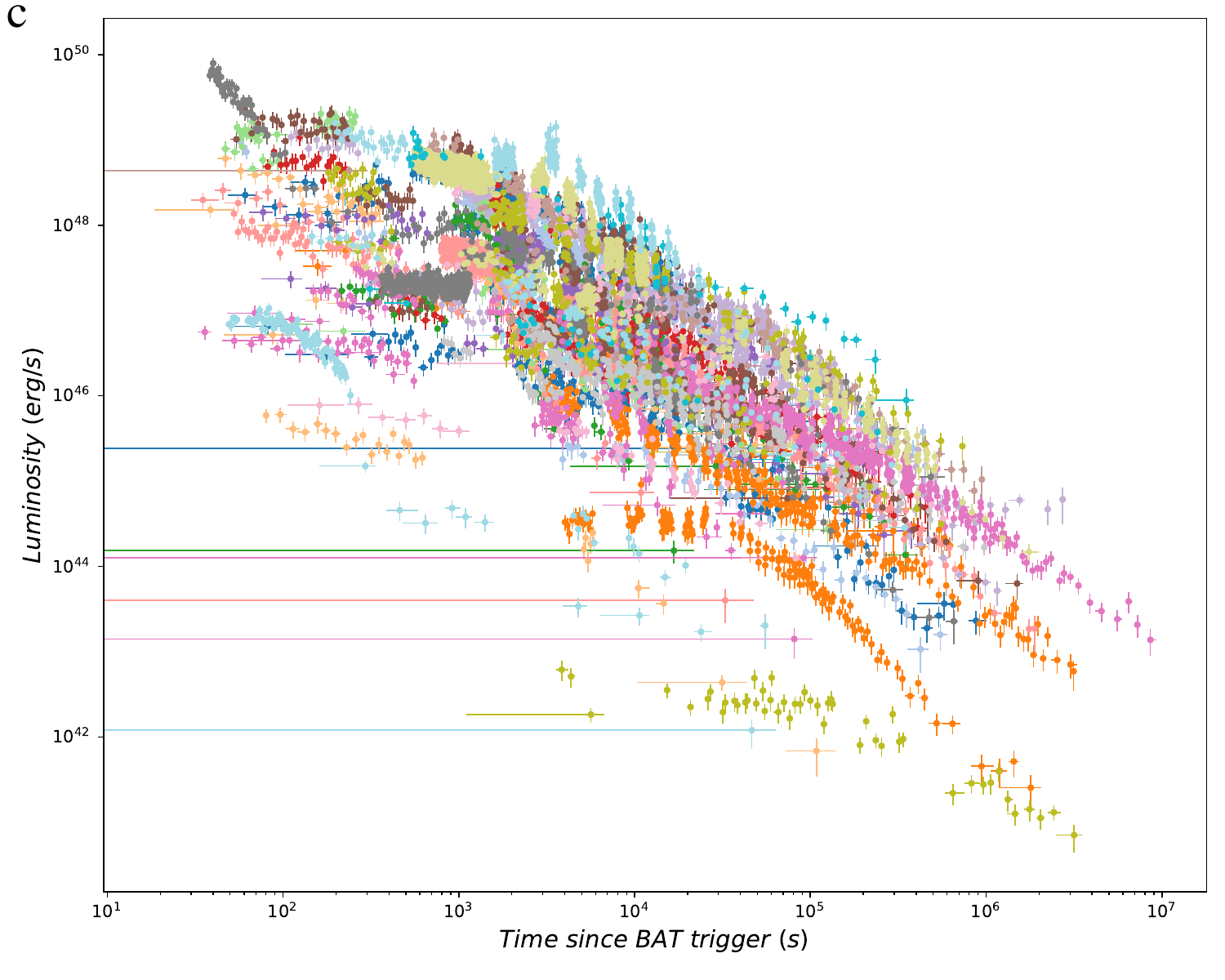}
	\includegraphics[width=0.45\columnwidth]{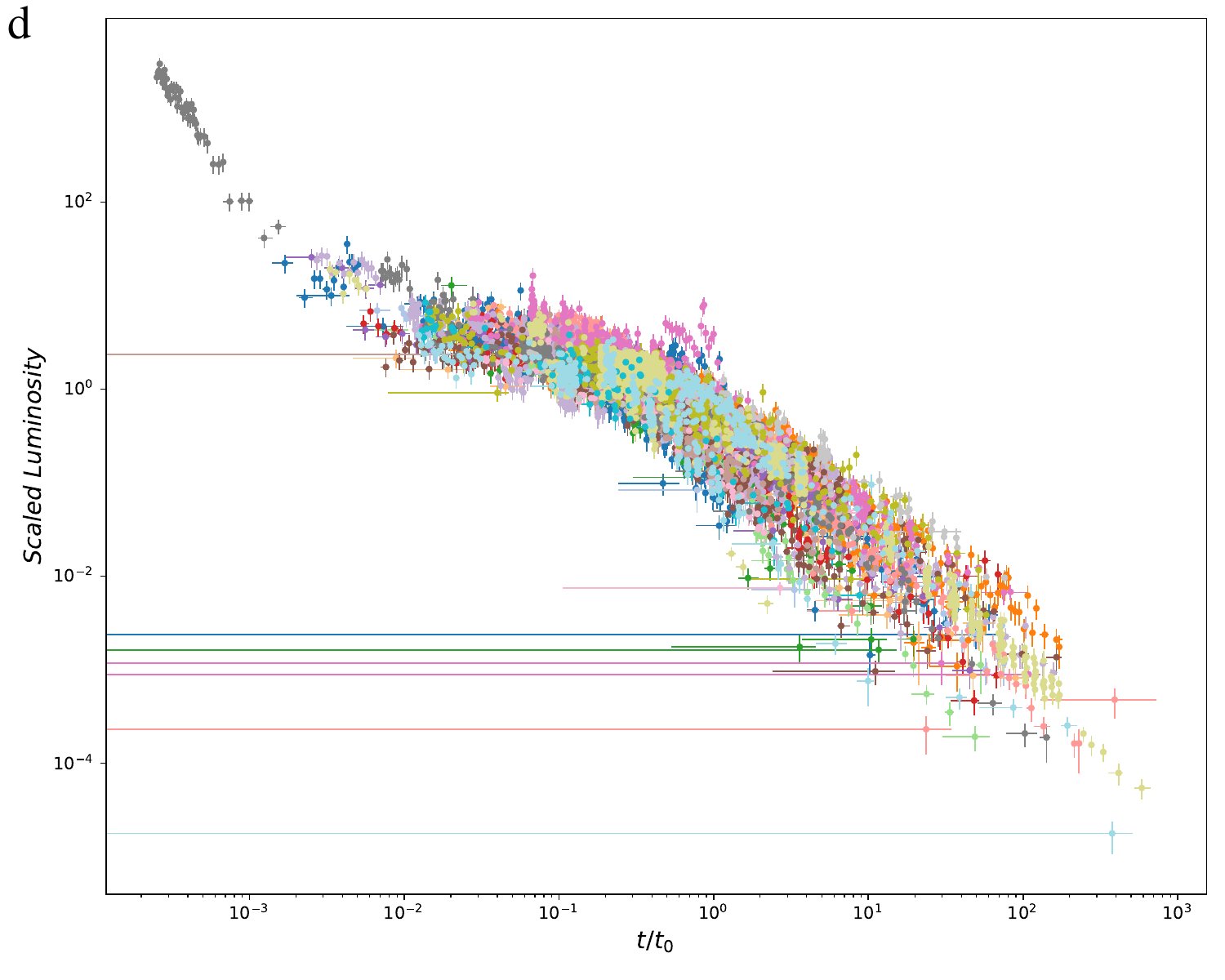}
	\includegraphics[width=0.45\columnwidth]{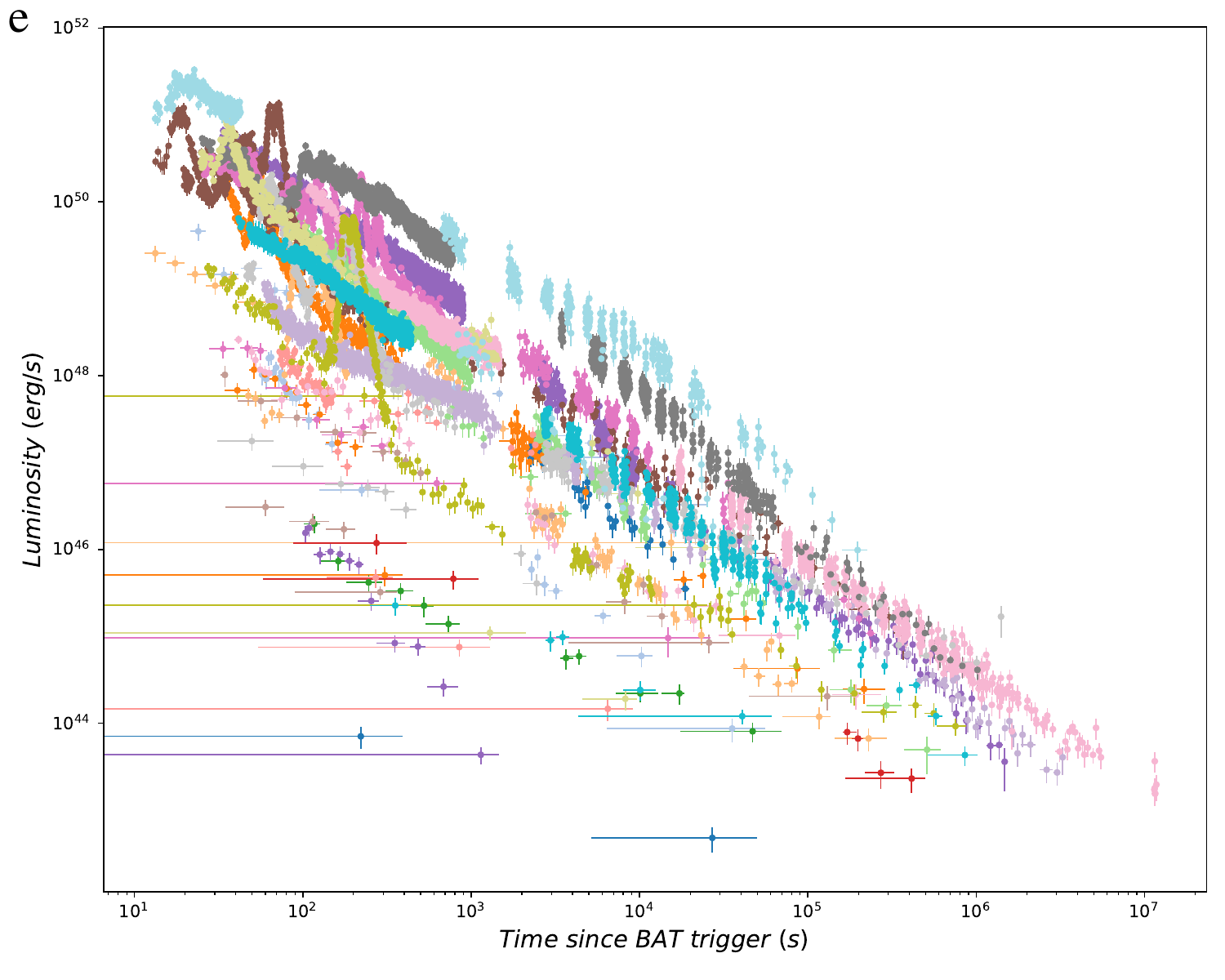}
	\includegraphics[width=0.45\columnwidth]{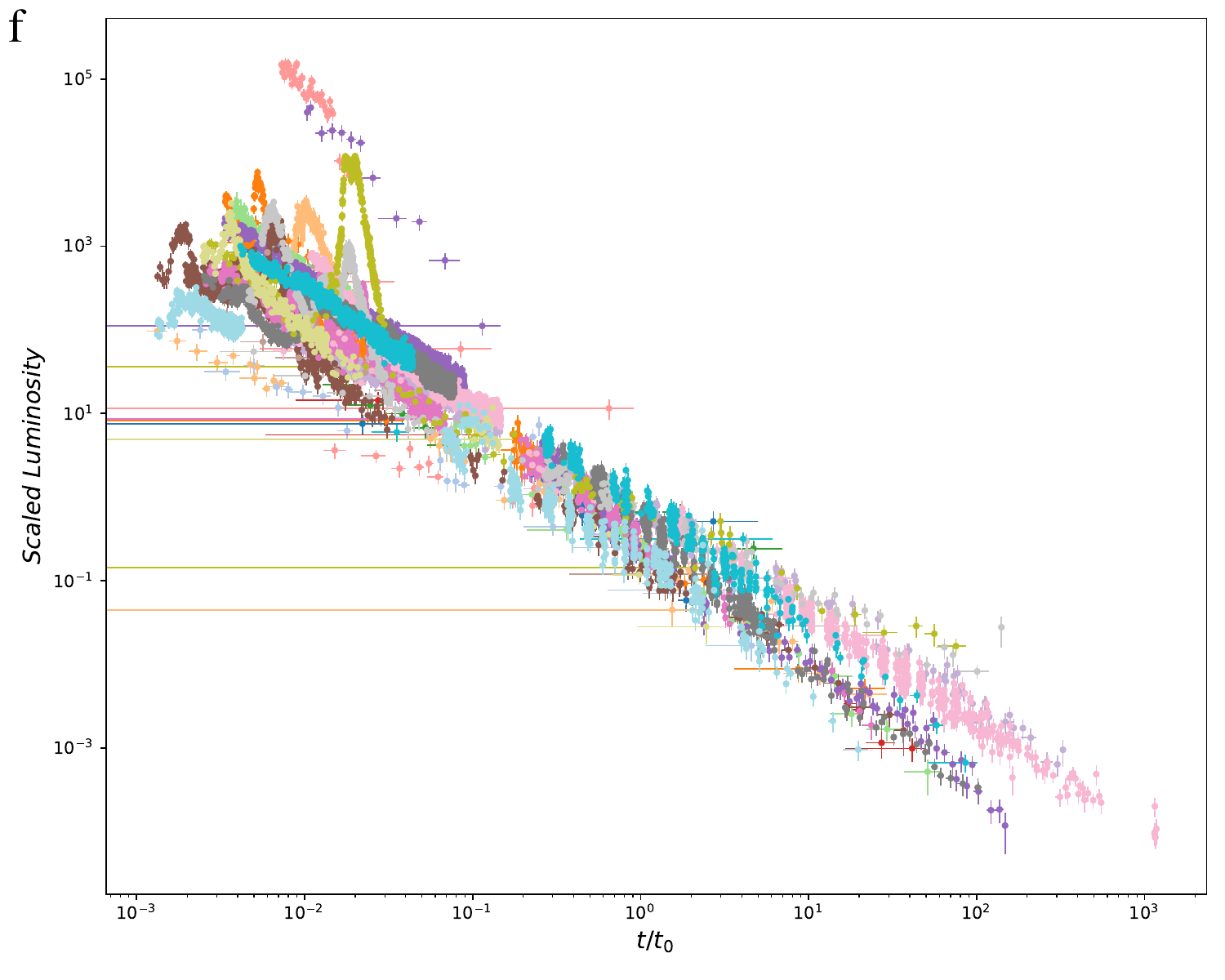}	
	\caption{The observed (left panel) and standardized (right panel) X-ray light curves with IXPs (top), EXPs (middle) and IFs (bottom), respectively.   }
	\label{fig:5}
\end{figure}

As shown in Section 3, a predominant fraction of 109/146 GRB X-ray afterglows can be successfully fitted by Eq. (\ref{4}), the remaining X-ray light curves can be simply fitted by the SPL form. Here, we will investigate how different are these GRBs with diverse observational properties or physical origins. In addition, each GRB will be examined separately in both the rest and the observer frames. We utilize Eq. (\ref{11}) to compute the isotropic X-ray luminosity of each GRB. For IXP and EXP samples, the light curve of GRB was normalized by selecting the plateau time of each GRB and the corresponding plateau photometricity, and for IF samples, the classical plateau time $t=10000$s and the photometric of the corresponding plateau were used to normalize the photometricity. The normalized image is shown in Figure \ref{fig:5}.

It is likely that accretion takes place in the collapsar origin magnetar-driven GRBs and probably produces some bumps in the afterglow (see e.g. Fraija et al. (2021)), however, no significant long-lasting accretion should occur in the case of SGRBs. Moreover, the duration of the plateau is typically $10^2-10^6$ seconds. Thus, if an accretion rate of $10^{-4}-10^{-1}$ solar masses per second would be maintained, the NS would easily grow to a mass bigger than 3 solar masses, therefore crossing the TOV limit and collapsing to a BH. In order to maintain a power-law-like slope after the end of the plateau, such a scenario in case of IF has to be ruled out.

\section{Distributional properties of parameters }
\subsection{Reshift distribution}
\begin{figure}
	\centering
	\includegraphics[width=8cm]{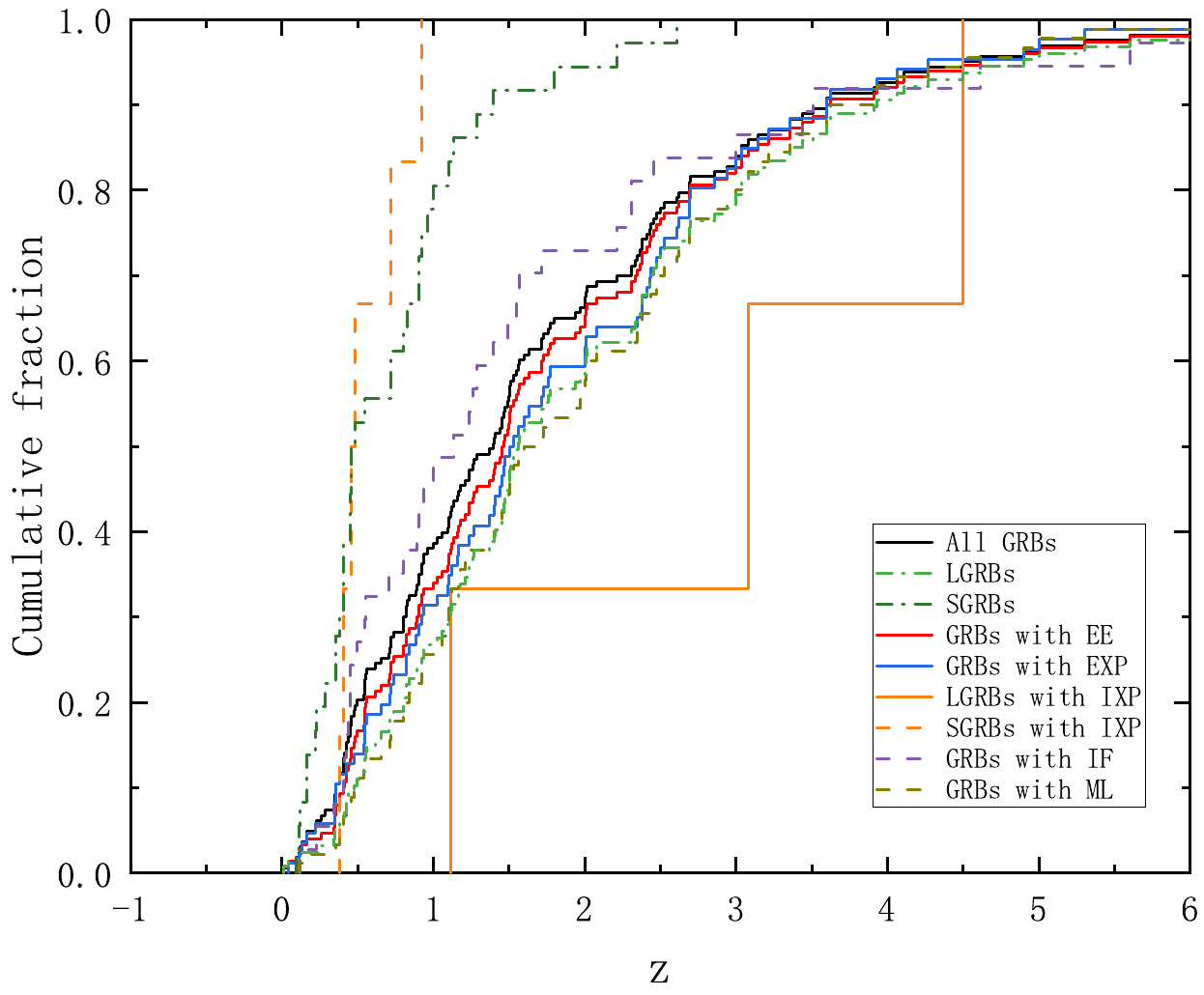}
	\includegraphics[width=8CM]{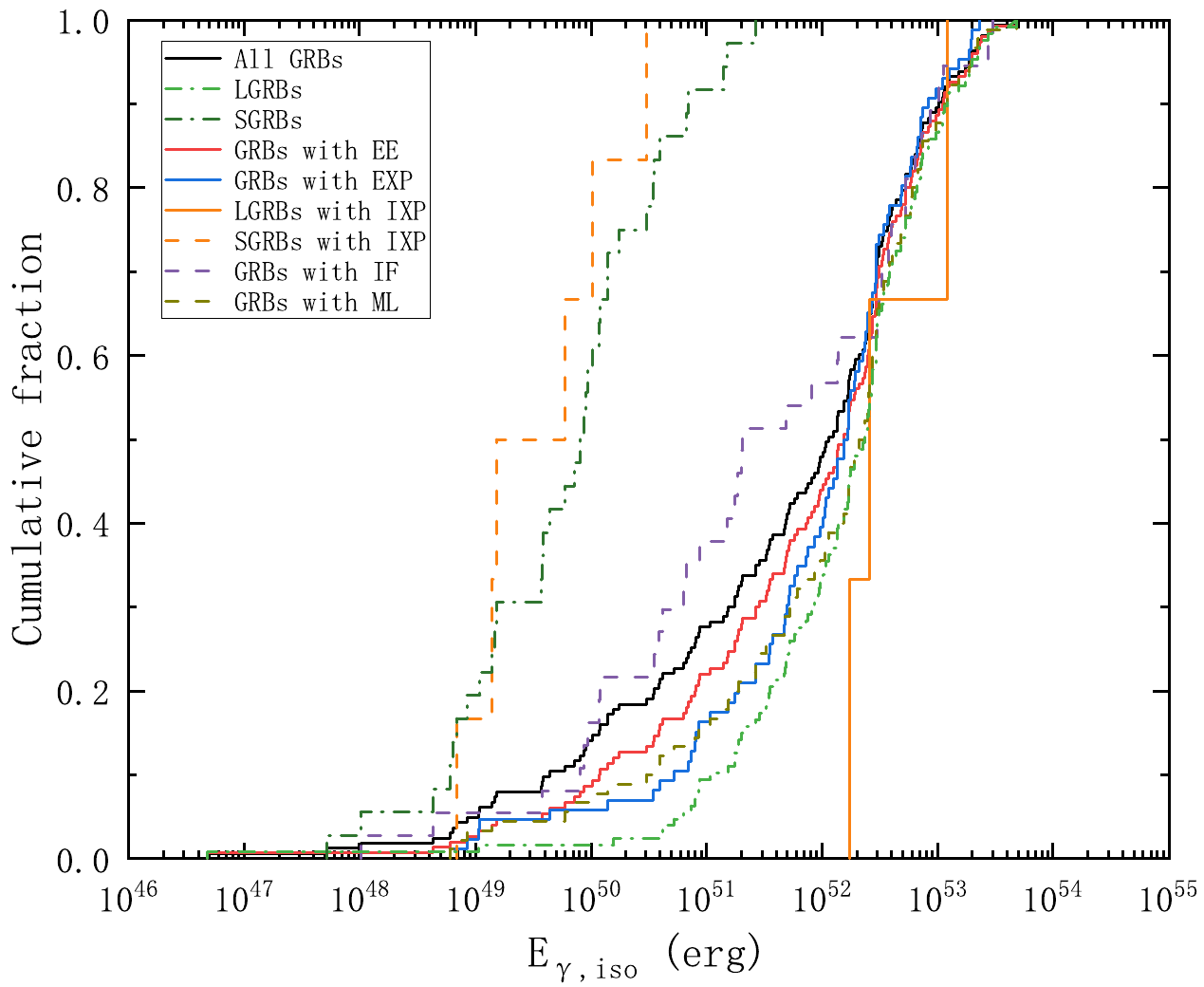}
	\caption{Cumulative redshift (left panel) and energy (right panel) distributions of different types of GRBs in our sample. The black solid line stands for the $z$ distribution of 163 GRBs in total. All symbols have been interpreted in the illustraion of each panel.}
	\label{fig:4}
\end{figure}

We plot the cumulative $z$ distributions of different types of GRBs in our sample in the left panel of Figure \ref{fig:4}, among which these GRBs have been also sorted into SGRBs and LGRBs besides the above X-ray based classes. Using the Kolmogorov-Smirnov (K-S) test to the LGRB and SGRB samples, we get the statistic $P={4.39\times10^{-8}}$ with a significance level of $\alpha=0.01$, indicating that LGRBs and SGRBs are different distributed. In the K-S test of redshift, the SGRB sample is only the same as the IXP and IF sample, and the distribution of other sub-samples is different. All testing results of any two $z$ distributions are presented in detail in Figure \ref{fig:7}, from which we notice that the IXPs have lower redshifts than the EXPs. Furthermore, we find that the IXP $z$ distribution is close to that of SGRBs, while the $z$ distributions of EXPs and LGRBs are much analogous. Other catagories including IF, ML and EE GRBs reside between SGRBs and LGRBs. The mean values of redshift across different samples are individually $\bar{z}_{EE}$= 1.80, $\bar{z}_{IXP}$= 1.34, $\bar{z}_{EXP}$= 1.83, $\bar{z}_{ML}$= 1.97, $\bar{z}_{IF}$= 1.59, $\bar{z}_{SGRB}$= 0.72 and $\bar{z}_{LGRB}$= 1.99. The mean redshift of the EE sample was comparable to that of the EXP, ML and LGRB samples.

\subsection{Energy distribution}

\begin{figure}
	\centering
	\includegraphics[width=8cm]{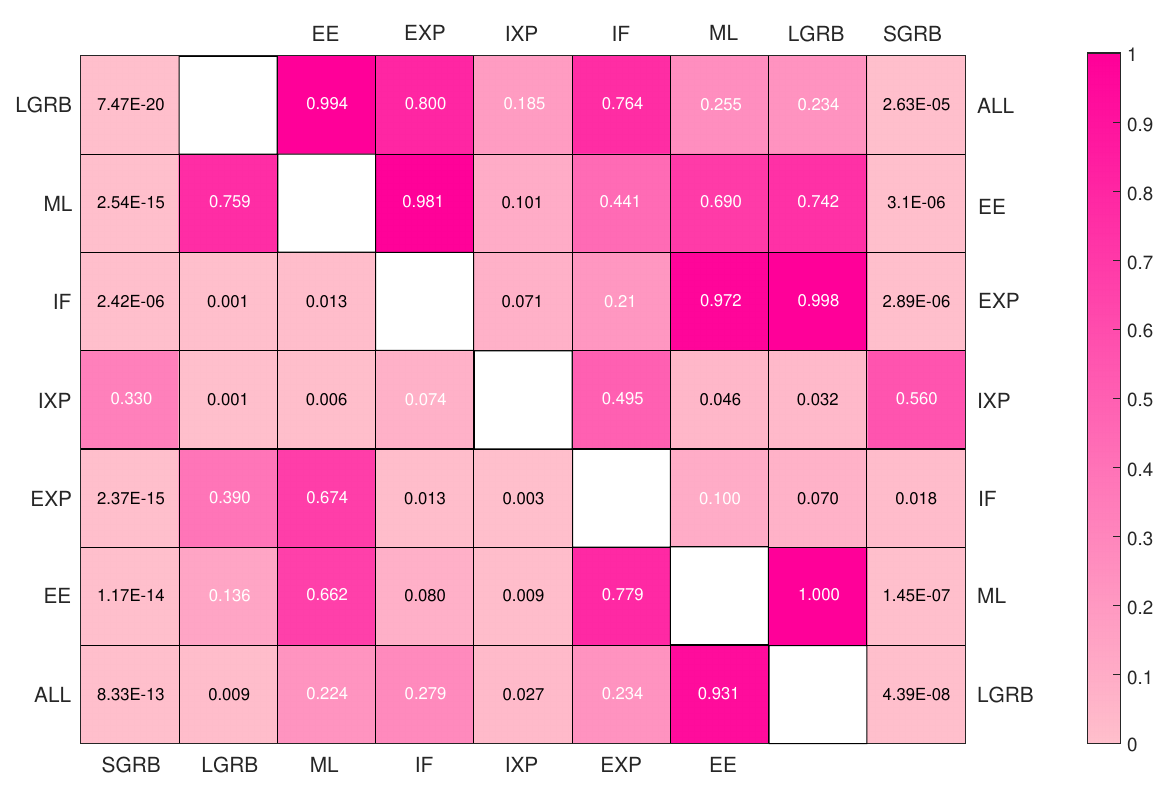}
	\includegraphics[width=8cm]{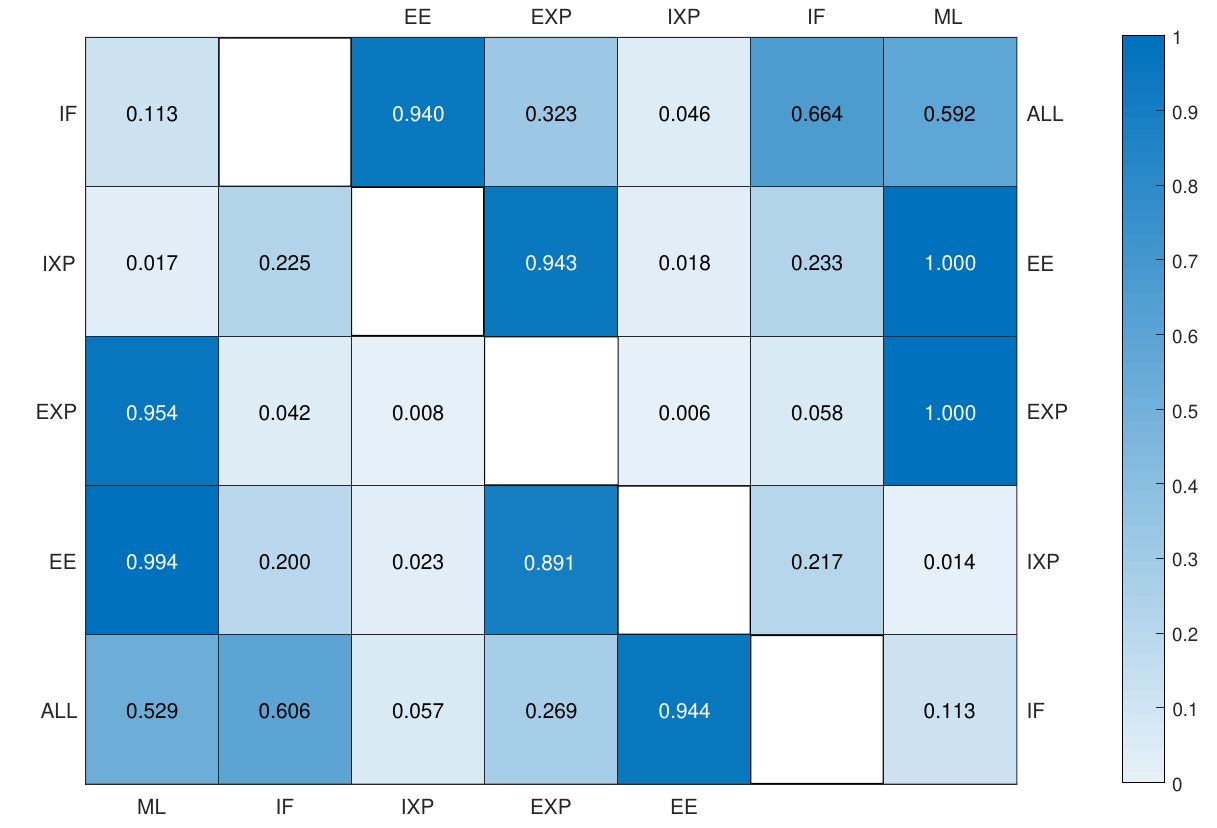}
	\caption{This is the result of the K-S test  ($\alpha=0.01$). The left panel shows the results of z and ${E_{\gamma,iso}}$, with the upper right triangle representing z and the lower left triangle representing ${E_{\gamma,iso}}$; the right panel shows the results of $T_{90}$ and $t_{\rm int}$, with the upper right triangle representing $T_{90}$ and the lower left triangle representing $t_{\rm int}$. }
	\label{fig:7}
\end{figure}

We compare the cumulative $E_{\gamma,iso}$ distributions of distinct kinds of GRBs in the right panel of Figure \ref{fig:4}, in which the distributional patterns are consistent with those of redshift in order. The IXPs/EXPs have the lowest/largest $E_{\gamma,iso}$ value on average. The median $E_{\gamma,iso}$ of LGRBs is about two orders of magnitude larger than that of SGRBs, which is consistent with the result of normal short and long GRBs \citep{2009ApJ...703.1696Z}. Interestingly, the distributions of LGRBs and SGRBs resemble those of EXPs and IXPs, respectively. The mean energies in different samples are individually ${\overline E _{\gamma ,iso}}\left( {EE} \right) = 3.92\times10^{52}$ erg, ${\overline E _{\gamma ,iso}}\left( {EXP} \right) = 3.27\times10^{52}$ erg, ${\overline E _{\gamma ,iso}}\left( {IXP} \right) = 1.85\times10^{52}$ erg, ${\overline E _{\gamma ,iso}}\left( {IF} \right) = 3.62\times10^{52}$ erg, ${\overline E _{\gamma ,iso}}\left( {ML} \right) = 4.35\times10^{52}$ erg, ${\overline E _{\gamma ,iso}}\left( {LGRB} \right) = 4.63\times10^{52}$ erg, ${\overline E _{\gamma ,iso}}\left( {SGRB} \right) = 2.74\times10^{50}$ erg. The K-S test results in Figure \ref{fig:7} indicate that the $E_{\gamma,iso}$ distributions of SGRB sample was only the same as the IXP sample.
\begin{figure*}
	\centering
	\includegraphics[width=8cm]{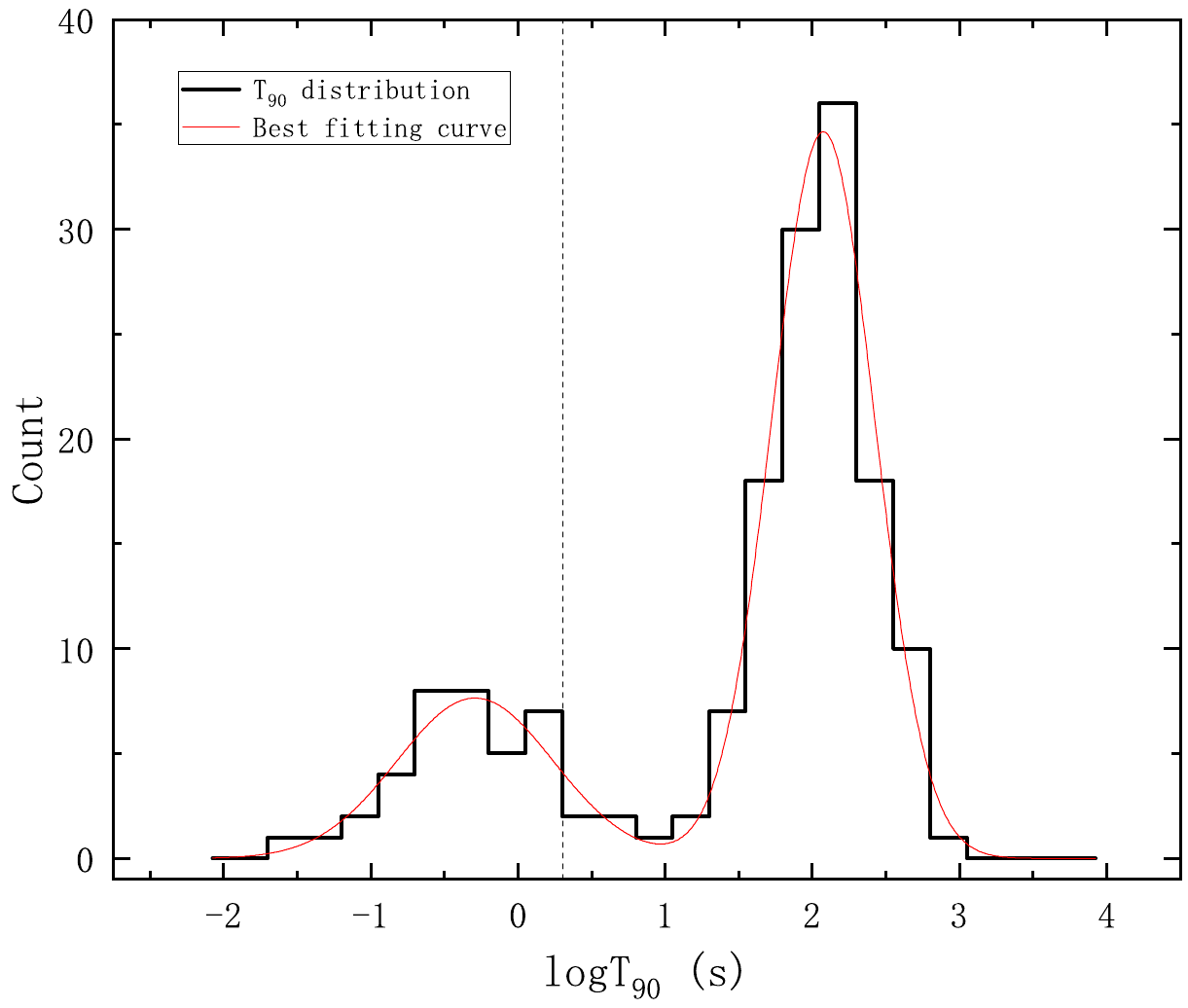}
	\includegraphics[width=8cm]{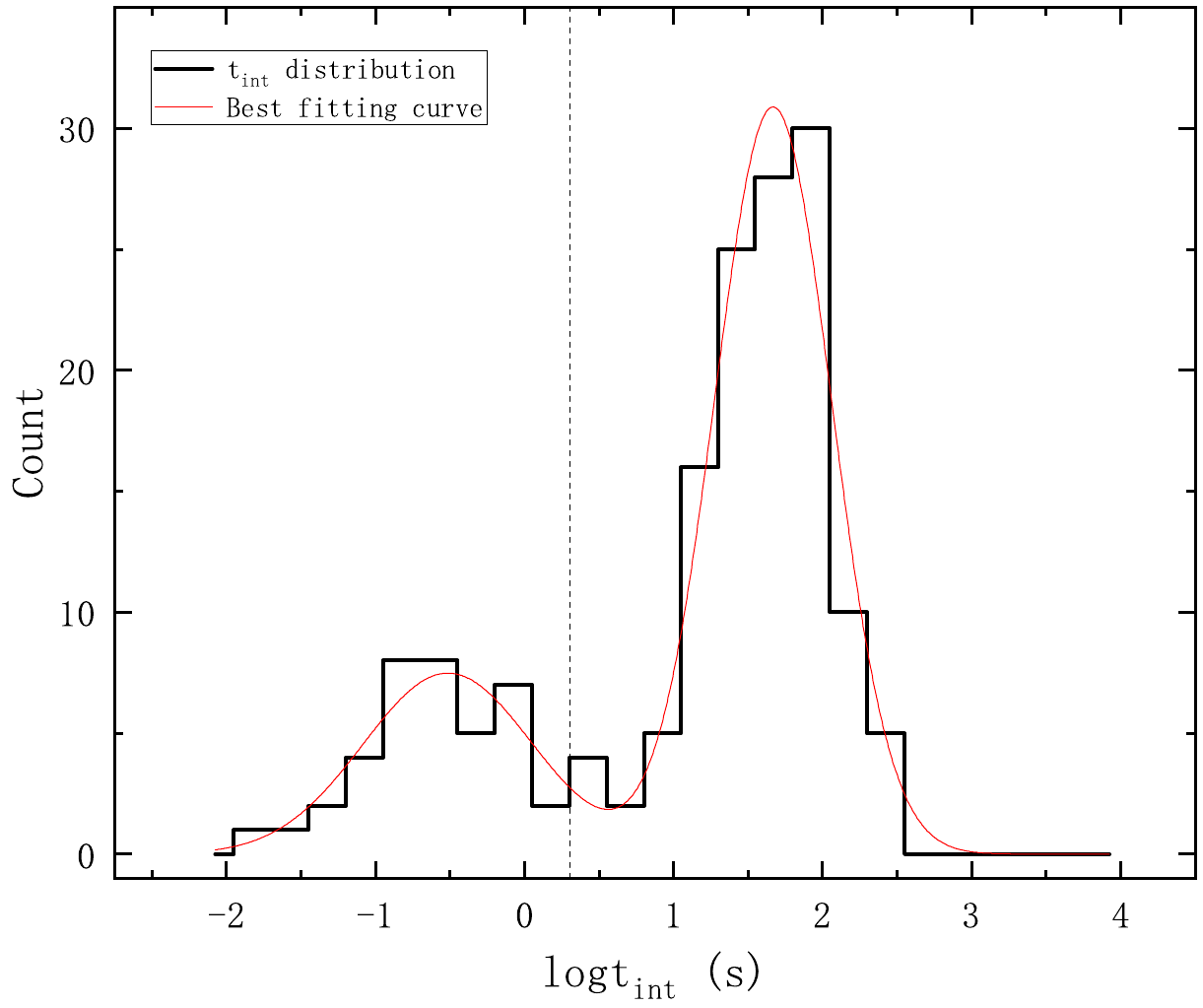}
	\caption{Distributional histograms of Swift GRBs are displayed in the observer (left panel) and rest (right panel) frames. The red solid curves stand for the best fit with a two-Gauss function. The vertical line denotes a duration of 2 seconds. }
	\label{fig:12}
\end{figure*}

\begin{figure}
	\centering
	\includegraphics[width=0.45\columnwidth]{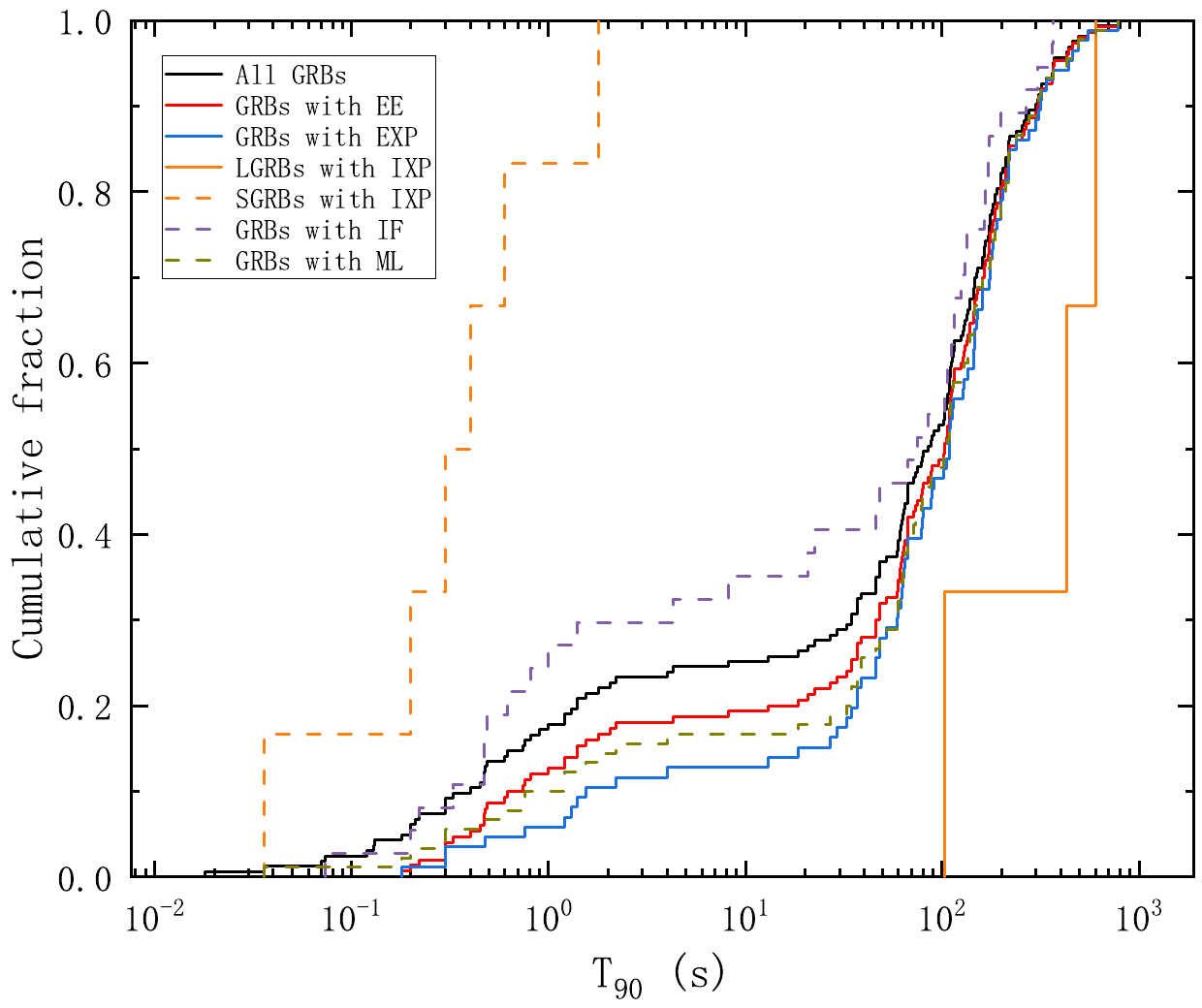}
	\includegraphics[width=0.45\columnwidth]{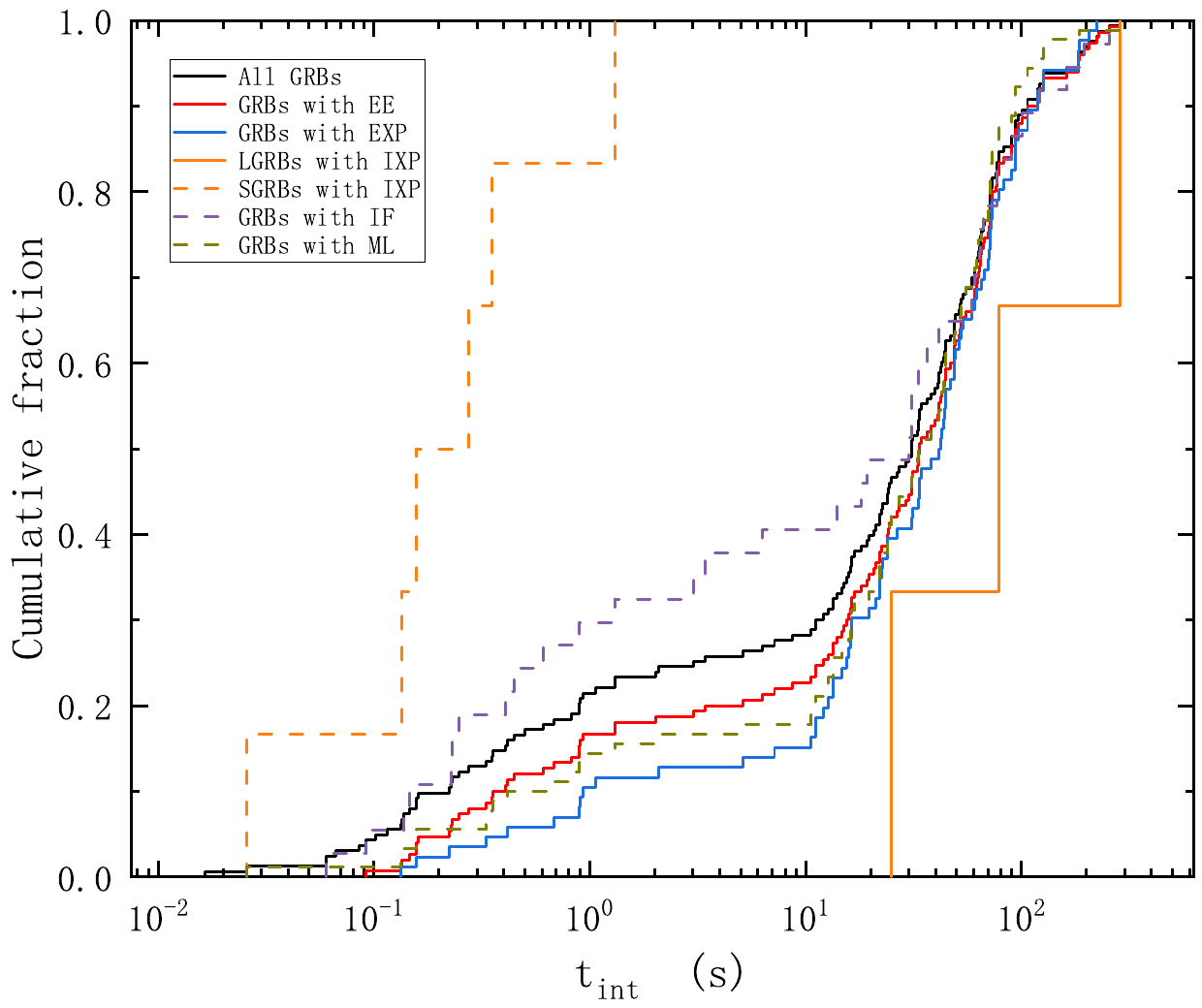}		
	\caption{Cumulative duration distributions of different kinds of GRBs in the observer frame on the left panel and in the rest frame on the right panel. All symbols are illustrated in the insert. The vertical line shows the boundary of 2 seconds.  }
	\label{fig:6}
\end{figure}
\subsection{Duration distribution}

Additionally, we make the distribution histograms of duration time in both observer and rest frames in Figure \ref{fig:12} and perform a two-Gauss fit to these histograms. It is verified with a K-S test that these distributions across distinct energy channels are identical with that in the total energy bands. The EE bursts with duration larger than 2s in our sample reside between SGRBs and LGRBs, which can confuse the boundary of two classes as suggested by \cite{2016MNRAS.462.3243Z}. Regardless of the existence of the middle class dominated by the EEs, the hardness ratios of the EEs and the LGRBs are equivalent in statistics, supporting that the dichotomy of \textit{Swift} GRBs is still tenable \citep{2022ApJ...940....5D}.


To examine the differences of duration distributions among different types of GRBs, we plot the cumulative distributions of durations in Figure \ref{fig:5}, where it is found that the IXPs are shorter than others in both observer and rest frames, while the EXPs hold the longest duration on average. Note that the EE sample comprises 125 LGRBs and 25 SGRBs, the IXP sample includes 6 SGRBs  and 3 LGRBs, the EXP sample contains 77 LGRBs and 9 SGRBs, the ML samples involves77 LGRBs and 13 SGRBs, and the IF sample consists of 26 LGRBs and 11 SGRBs. We perform a K-S test to compare any two of these distributions and list the results in Figure \ref{fig:7}. The K-S test of $T_{90}$ showed that only the IXP sample had different distributions from EXP samples. The K-S test of $t_{int}$ also showed that only the IXP sample had different distributions from EXP samples.

\section{Summary}

Based on the above investigations, we can obtain the strong contraints on diverse kinds of GRBs most probabaly driven by magnetars. For this purpose, we focus on these special GRBs with X-ray plateaus, power-law decay afterglows and/or extended emissions in $\gamma$-rays. By analyzing some empirical relations and parameter distributions, we obtain the following important conclusions:
\begin{itemize}
	\item[1)] 
	In our sample, 150 GRBs with EEs, 86 GRBs with EXPs, 9 GRBs with IXPs, 14 GRBs with X-ray quasi-plateaus and 37 GRBs with IF X-ray afterglows have been successfully distinguished. 
	\item[2)]   
	Using 109 GRBs with X-ray plateaus or quasi-plateaus, we build the three-parameter relation of $L_X\varpropto T_a^{-1.13}E_{\gamma,iso}^{0.74}$, of which the temporal index is extremely consistent with previous works \cite[e.g.][]{2012A&A...538A.134X,2019ApJS..245....1T} and the energy index of 0.74 is marginally consistent with their value of $\sim$0.8 within a 2-$ \sigma$ level, supporting that our sub-sample of GRBs are most likely powered by the magnetars.
	\item[3)] 
	Although GRBs with IXPs are highly proposed to originate from magnetars, we testify in the $B-P$ plot that those GRBs with either EXPs, X-ray quasi-plateaus or $\gamma$-ray EEs could also be driven by the central engine of magnetars.
	\item[4)] 
	 It is found that the averaged X-ray light curves of GRBs with EEs are very similar to those of GRBs with EXPs or the ML sample in both the observer and the rest frames, which strongly convince us that these GRBs with EXPs, IXPs and/or EEs should be produced from magnetars. In contrast, the fainter IXPs break earlier than the EXPs as a whole. 
	  \item[5)] 
	  The K-S test results of z showed that the SGRB sample was only the same as the IXP and IF sample. In the K-S test of $E_{\gamma,iso}$, the SGRB sample is only the same as the IXP sample, which is different from the distribution of other sub-samples, the ML sample is different from the distribution of IXP. 
	  From the results of the K-S test of $T_{90}$, the EXP sample had different distributions from IXP sample. From the results of the K-S test of $T_{int}$, the EXP samples were different from the IXP sample.
	 
 
\end{itemize}
  
%
%
%

\bibliography{sample631}{}
\bibliographystyle{aasjournal}

\section*{Acknowledgments}
This work was supported by National Natural Science
Foundation of China (grant Nos. U2031118 and 62105097). We acknowledge the usage of the archive data on the\textit{ Swift} website (https://swift.gsfc.nasa.gov/archive/grb\_table.html/).
\begin{longrotatetable}
\begin{deluxetable*}{lccccccccc}
		\tabletypesize{	\scriptsize}
		\renewcommand{\arraystretch}{0.6} 
		\label{tab:1}
		\tablecaption{Basic $\gamma$-ray data of \textbf{163} GRBs with multi-segments in our sample}
		\tablehead{
			\colhead{GRB}
			&\colhead{$z$}
			&\colhead{$T_{90}$ (s)}
			&\colhead{Precursor}
			&\colhead{Main peak}
			&\colhead{EEs}
			&\colhead{X-ray type}
			&\colhead{$S_{\gamma}$ ($\times 10^{-7} \rm erg\ \rm cm^{-2}$)}
			&\colhead{ ${K_{c}}$ }
			&\colhead{${E_{\gamma,iso}}$ (erg)}\\
			(1)&(2) & (3)& (4)& (5)& (6)&(7) &(8) &   (9)   &(10)
		}
		\startdata
		050505  & 4.27     & 58.9              & Y   & 1  & Y  & EXP,ML & 24.9$\pm$1.79     & 0.38 & (3.41$\pm$0.24)$\times {\rm{1}}{{\rm{0}}^{{\rm{52}}}}$ \\
		050509B & 0.225    & 0.07              & Y   & 1  & N  & IF     & 0.09$\pm$0.02     & 0.92 & (1.02$\pm$0.23)$\times {\rm{1}}{{\rm{0}}^{{\rm{48}}}}$ \\
		050709  & 0.1606   & 0.07$^a$              &     &    & N  & ...    & 2.90$\pm$0.40$^b$     & 0.82 & (1.47$\pm$0.20)$\times {\rm{1}}{{\rm{0}}^{{\rm{49}}}}$ \\
		050724  & 0.257    & 96                & Y   & 1  & Y  & ...    & 9.98$\pm$1.20     & 0.98 & (1.59$\pm$0.19)$\times {\rm{1}}{{\rm{0}}^{{\rm{50}}}}$ \\
		050803  & 0.422    & 87.9              & Y   & 2  & Y  & EXP,ML & 21.50$\pm$1.35    & 0.80 & (7.94$\pm$0.50)$\times {\rm{1}}{{\rm{0}}^{{\rm{50}}}}$ \\
		050813  & 1.8      & 0.45              & N   & 1  & Y  & ...    & 0.44$\pm$0.11     & 0.48 & (1.76$\pm$0.44)$\times {\rm{1}}{{\rm{0}}^{{\rm{50}}}}$ \\
		050814  & 5.3      & 150.9             & Y   & 2  & Y  & EXP,ML & 20.10$\pm$2.20    & 0.69 & (7.07$\pm$0.77)$ \times {\rm{1}}{{\rm{0}}^{{\rm{52}}}}$ \\
		050915A & 2.5273   & 52                & Y   & 2  & Y  & EXP,ML & 8.50$\pm$0.88     & 0.46 & (6.06$\pm$0.63)$\times {\rm{1}}{{\rm{0}}^{{\rm{51}}}}$ \\
\enddata
		\tablenotetext{a} {The $T_{90}$ value from \cite{2020ApJ...895...58G};}
		\tablenotetext{b} {The fluence value from \cite{2005GCN..3385....1B};}
		\tablenotetext{c} {The fluence value from \cite{2017GCN.21520....1V};}
		\tablenotetext{d}{The ${K_{c}}$ value from \cite{2018PASP..130e4202Z};}
		Note: 1.z values, $T_{90}$ values and $S_{\gamma}$ values are from \url{ https://www.mpe.mpg.de/~jcg/grbgen.html}; \url{https://swift.gsfc.nasa.gov/results/batgrbcat/} and \url{https://swift.gsfc.nasa.gov/archive/grb_table/index.php};
		2."EE"="Extended emission","IF"="Power-law","ML"="Magnetar-like","IXP"="Internal X-ray plateau","EXP"="External X-ray plateau";
		3.${E_{\gamma,iso}}$ calculated from equation \ref{1}
	\end{deluxetable*}
\end{longrotatetable}

{
	\tiny  
	\footnotesize 
	\renewcommand{\arraystretch}{0.65} 
	\begin{longtable}{cccccc}
		
		\caption{Best fitting parameters of 146 GRB X-ray afterglows}\label{tab:2} \\
		
		\hline
		GRB & $\alpha_1$ & $\alpha_2$ &$T_0$ (s) &{$F_{\rm x0}$} ($\rm erg\ \rm cm^{-2}\ \rm s^{-1}$)&Type  \\
		\hline
		\endfirsthead 
		\hline
		GRB & $\alpha_1$ & $\alpha_2$ &$T_0$ (s) &{$F_{\rm x0}$} ($\rm erg\ \rm cm^{-2}\ \rm s^{-1}$)&Type  \\
		\hline
		\endhead 
		\midrule
		\endfoot 
		\bottomrule
		\endlastfoot 
		
		050505  & 0.60$\pm$0.05  & 1.67$\pm$0.05        & 19539.42$\pm$2290.95   & (8.73$\pm$1.22)$ \times {\rm{1}}{{\rm{0}}^{{\rm{-12}}}}$   & EXP,EE \\
		050509B & -0.56$\pm$0.00 & \textbackslash{} & \textbackslash{}   & \textbackslash{}                     & IF     \\
		050803  & 0.47$\pm$0.06  & 1.59$\pm$0.07        & 14585.23$\pm$2648.71   & (1.12$\pm$0.20)$ \times {\rm{1}}{{\rm{0}}^{{\rm{-11}}}}$  & EXP,EE \\
		050814  & 0.62$\pm$0.03  & 2.10$\pm$0.25        & 78890.86$\pm$14103.17  & (5.94$\pm$1.12)$ \times {\rm{1}}{{\rm{0}}^{{\rm{-13}}}}$  & EXP,EE \\
		050915A & 0.37$\pm$0.37  & 1.07$\pm$0.30        & 917.87$\pm$1546.13     & (1.79$\pm$2)$ \times {\rm{1}}{{\rm{0}}^{{\rm{-11}}}}$     & EXP,EE \\
		051006  & 0.80$\pm$0.82  & 1.94$\pm$1.47        & 647.52$\pm$1101.98     & (4.01$\pm$8.72)$ \times {\rm{1}}{{\rm{0}}^{{\rm{-11}}}}$  & EE     \\
		051016B & 0.41$\pm$0.04  & 1.32$\pm$0.06        & 24219.57$\pm$4530.46   & (1.92$\pm$0.31)$ \times {\rm{1}}{{\rm{0}}^{{\rm{-12}}}}$   & EXP    \\
		051109A & 0.48$\pm$0.04  & 1.26$\pm$0.02        & 5982.11$\pm$978.71     & (3.29$\pm$0.53)$ \times {\rm{1}}{{\rm{0}}^{{\rm{-11}}}}$  & EXP,EE \\
\end{longtable}
}

\begin{longrotatetable}
	\begin{deluxetable*}{lccccccccc}
		\renewcommand{\arraystretch}{0.6} 
		\label{tab:3}
		\tablecaption{The $L-T-E$ correlation parameters and the best $B-P$ fitting parameters of \textbf{109} GRBs with X-ray plateaus including IXPs, EXPs and others.}
		\tablehead{
			\colhead{GRB}
			&\colhead{$log \left( {{T_{\rm{a}}}/{{10}^3\ \rm s}} \right)$ }
			&\colhead{ $log \left( {{L_X}/{{10}^{47}}\ \rm erg\ s^{-1}} \right)$ }
			&\colhead{$log \left( {E_{\gamma ,iso}}/{{10}^{53}}\rm erg\right)$ }
			&\colhead{$B$ ($\times{10}^{14} \rm G$) }
			&\colhead{ $P$ (ms)}
			&\colhead{Afterglow type}
			&\colhead{ML candidate$^{\star}$  }
			&\colhead{ $\Gamma $ }
			&\colhead{${\theta_j}^{\ast}$ (deg)} \\
			(1)&(2) & (3)& (4)& (5)& (6)&(7) &(8) &   (9)   &(10)
		}
		\startdata
		050505  & 0.57$\pm$0.05                                         & 1.24$\pm$0.05                        & -0.47$\pm$0.08 & 9.92$\pm$0.17       & 1.58$\pm$0.01     & EXP              & Y  & 2.02                       & $2.10^g$  \\
		050803  & 1.01$\pm$0.07                                         & -1.12$\pm$0.07                       & -2.10$\pm$0.03 & 73.22$\pm$3.73      & 20.94$\pm$0.49    & EXP              & Y  & 2.12                       & $5.00$   \\
		050814  & 1.10$\pm$0.07                                         & 0.26$\pm$0.07                        & -0.15$\pm$0.06  & 15.61$\pm$1.56      & 2.27$\pm$0.09     & EXP              & Y  & 1.97                       & $2.69^f$  \\
		050915A & -0.58$\pm$0.27                                        & 1.10$\pm$0.23                        & -1.22$\pm$0.09 & 111.57$\pm$16.03    & 8.83$\pm$0.55     & EXP              & Y  & 2.21                       & $13.75^f$ \\
		051006  & -0.50$\pm$0.27                                        & 0.48$\pm$0.30                        & -1.54$\pm$0.06 & 318.42$\pm$79.48    & 12.73$\pm$0.59    & \textbackslash{} & Y  & 2.24                       & $16.62^f$ \\
		051016B & 1.10$\pm$0.07                                         & -1.08$\pm$0.06                       & -2.28$\pm$0.04 & 57.12$\pm$3.18      & 21.92$\pm$0.53    & EXP              & Y  & 1.89                       & $5.40^g$  \\
		051109A & 0.25$\pm$0.06                                         & 1.21$\pm$0.06                        & -0.78$\pm$0.09 & 14.48$\pm$0.57      & 1.95$\pm$0.02     & EXP              & Y  & 2.07                       & $3.40^g$  \\
		051221A & 1.26$\pm$0.07                                         & -1.76$\pm$0.07                       & -2.16$\pm$0.02 & 1941.73$\pm$181.48  & 998.79$\pm$34.51  & EXP              & N  & 2.06                       & $12.61^a$ \\
\enddata
	\textbf{Note:}\\ 1. Symbol Y indicates that the GRB is a magnetar-like candidate under the B-P limit, and Symbol N indicates that the GRB is not a magnetar-like candidate.\\ 2. References for $\theta_j$, namely $^a${\cite{2024MNRAS.530.2877L};}	  $^b${\cite{2021ApJ...908L...2S};} $^c${\cite{2023ApJ...959...13R};} $^d${\cite{2023ApJ...949L..32D};} $^e${\cite{2023ApJ...950...30Z};} $^f${\cite{2015ApJ...799....3R};} $^g$ {\cite{2018PASP..130e4202Z}.}\\ 3. The $\theta_j$ values are assumed for SGRBs as $\theta_j=10^{\circ}$ and LGRBs as $\theta_j=5^{\circ}$, respectively, if the jet angles had not been reported, since the averaged jet angle of LGRBs is about $5^{\circ}$ that is slightly smaller than that of SGRBs statistically \citep{2006ApJ...650..261S}.

		\end{deluxetable*}
	\end{longrotatetable}
\newpage

\end{document}